\documentclass[aps,prx,groupedaddress,showpacs,twocolumn]{revtex4-1}
\usepackage[utf8]{inputenc}
\usepackage{hyperref}
\usepackage{graphicx}  
\usepackage{dcolumn}   
\usepackage{bm}        
\usepackage{amssymb,amsmath}   
\usepackage[normalem]{ulem}
\usepackage{uri}

\usepackage{color}

\usepackage[x11names, rgb, html]{xcolor}
\definecolor{sbase03}{HTML}{002B36}
\definecolor{sbase02}{HTML}{073642}
\definecolor{sbase01}{HTML}{586E75}
\definecolor{sbase00}{HTML}{657B83}
\definecolor{sbase0}{HTML}{839496}
\definecolor{sbase1}{HTML}{93A1A1}
\definecolor{sbase2}{HTML}{EEE8D5}
\definecolor{sbase3}{HTML}{FDF6E3}
\definecolor{syellow}{HTML}{B58900}
\definecolor{sorange}{HTML}{CB4B16}
\definecolor{sred}{HTML}{DC322F}
\definecolor{smagenta}{HTML}{D33682}
\definecolor{sviolet}{HTML}{6C71C4}
\definecolor{sblue}{HTML}{268BD2}
\definecolor{scyan}{HTML}{2AA198}
\definecolor{sgreen}{HTML}{1C8909}
\hypersetup{
  colorlinks = true,
  allcolors = {sgreen}
}

\usepackage[T1]{fontenc}

\newcommand{\kb     }{k_{\rm B}}

\begin{document}

\title{ Quantifying entropy production in active fluctuations of the hair-cell bundle from time irreversibility and uncertainty relations}
\author{\'Edgar Rold\'an$^{1,2}$, J\'er\'emie Barral$^{3,4}$,  Pascal Martin$^{3,4}$, Juan M. R. Parrondo$^5$, and Frank J\"ulicher$^{2}$}
\affiliation{$^1$ICTP - The Abdus Salam International Centre for Theoretical Physics, Strada Costiera 11, 34151, Trieste, Italy \\
$^2$ Max Planck Institute for the Physics of Complex Systems, N{\"o}thnitzer Str. 38, 01187 Dresden, Germany \\
$^3$Laboratoire Physico-Chimie Curie, Institut Curie, PSL Research University, CNRS, UMR168, F-75248 Paris, France\\
$^4$Sorbonne Universit\'e, UPMC Univ Paris 06, F-75252 Paris, France\\
$^5$Departamento de Estructura de la Materia, F\'isica Termica y Electronica and GISC, Universidad Complutense de Madrid 28040 Madrid, Spain}

\begin{abstract}

We introduce  lower bounds for the rate of entropy production of an active stochastic process by quantifying the irreversibility of stochastic traces obtained from mesoscopic degrees of freedom.
Our measures of irreversibility  reveal signatures of time's arrow and provide bounds for entropy production even in the case of active fluctuations that have no drift.
  We apply these irreversibility measures to experimental recordings of spontaneous hair-bundle oscillations in mechanosensory hair cells from the ear of the bullfrog.  By analysing the fluctuations of only the tip position of hair bundles, we reveal irreversibility in active oscillations and estimate an associated rate of entropy production  of at least $\sim 3 k_{\rm B}$/s,  on average. Applying thermodynamic uncertainty relations, we predict that measuring  both the tip position  of the hair bundle and the mechano-electrical transduction current that enters the hair cell leads to tighter lower bounds for the rate of entropy production, up to $\sim 10^3 k_{\rm B}$/s in the oscillatory regime.
\end{abstract}

\maketitle

\section{Introduction}
\label{sec:1}
Active systems are maintained out of equilibrium by processes that consume resources of energy and  produce entropy.   This is the case of living cells, where energy is provided in the form of biochemical fuel such as adenosine triphosphate that drives active mesoscopic cellular processes. As discussed below, an important example of active cellular fluctuations  are spontaneous oscillations of mechanosensory hair bundles of auditory hair cells~\cite{martin2003spontaneous,tinevez2007unifying}. These oscillations have been proposed to amplify sound stimuli in the ear of many vertebrates, providing exquisite sensitivity and sharp frequency selectivity~\cite{hudspeth2014integrating}. 

Active mesoscopic processes 
do not obey the fluctuation-dissipation theorem: measuring both the linear response of the system to weak external stimuli and spontaneous  fluctuations  provides a means to quantify deviations from thermal equilibrium~\cite{martin2001comparison,harada2005equality,mizuno2007nonequilibrium,rodriguez2015direct,turlier2016equilibrium,battle2016broken,nardini2017entropy}. A related important question is how entropy production can be estimated in active mesoscopic systems.
In  cases where active systems generate movement with drift, such as molecular motors moving along  filaments~\cite{julicher1997modeling,keller2000mechanochemistry,howard2001mechanics},  
 the rate of entropy production can be estimated from measurements of drift velocities and  viscous forces~\cite{julicher1997modeling,qian2000mathematical}.
However, for active fluctuations without drift, such as spontaneous oscillations, 
it is unclear how entropy production  can be characterized.  Time irreversibility is a signature of the nonequilibrium nature of a system~\cite{steinberg1986time}. This suggests that quantification of irreversibility of fluctuations provides information about entropy production.

Hair cells are the cellular microphones of the inner ear~\cite{hudspeth1989ear}. They transduce sound-evoked mechanical vibrations of their hair bundle---a cohesive tuft of cylindrical stereocillia that protrudes from their apical surface (Fig.~\ref{fig:1}A)---into electrical signals that then travel to the brain. Fluctuations and response of the hair bundle provide a paradigmatic case study of nonequilibrium physics in biology.   Hair bundles from the ear of the bullfrog show noisy spontaneous oscillations ~\cite{martin2003spontaneous}. Under periodic external stimulation, oscillatory hair bundles can  actively amplify their response, resulting in a hysteretic behaviour corresponding to a net  energy extraction from the bundle~\cite{martin1999active}.  Furthermore, it was shown that  the fluctuation-dissipation theorem does not hold for oscillatory hair bundles, revealing that  their spontaneous fluctuations are also active~\cite{martin2001comparison}.  This finding demonstrates that hair bundle fluctuations must be described by   nonequilibrium stationary states.  Such behaviour can be captured by a minimal two-variable stochastic model with nonlinear and non-conservative forces~\cite{martin2003spontaneous,nadrowski2004active,le2005adaptive,tinevez2007unifying,martin2020mechanical}. In these models, one variable describes the tip position of the hair bundle, whereas the other variable describes the dynamics of a collection of  molecular  motors that power the bundle oscillations. Although the tip position can be measured,  motors' fluctuations are  hidden and hence can only be estimated  from stochastic simulations~\cite{nadrowski2004active,tinevez2007unifying,dinis2012fluctuation}. Experimental and theoretical evidence led to the proposal that hair bundle spontaneous fluctuations are akin to noisy limit-cycle oscillations close to a Hopf bifurcation~\cite{martin2020mechanical}.  Hence, as any active system,  hair-bundle spontaneous fluctuations are   characterized by probability fluxes in suitable phase spaces and by entropy production. Whether, and to what extent, tools from the emerging field of stochastic thermodynamics~\cite{sekimoto2010stochastic,seifert2008stochastic} can be used to estimate entropy production from measurements of active hair-bundle fluctuations remains an open question. 

 In this work, we introduce  and put to the test a hierarchy of bounds for the steady-state rate of entropy production based on measures of  irreversibility of sets of mesoscopic observables.
 We show that quantifying irreversibility can reveal  whether a noisy signal is produced by an active process or by a passive system.
 We apply the theory to experimental recordings of 
 spontaneous mechanical oscillations of mechanosensory hair bundles in an excised preparation from the ear of the bullfrog~\cite{barral2018friction}. Quantifying irreversibility from measurements of the bundle tip position,  we obtain lower bounds for the  entropy production of  its  spontaneous fluctuations. Finally, by means of uncertainty relations~\cite{barato2015thermodynamic,gingrich2016dissipation}, we  show that tighter bounds of entropy production can be obtained if one also measures the mechano-electrical  transduction current.
 
The paper is organized as follows. In Sec.~\ref{sec:2}, we discuss generic properties of irreversibility and dissipation of mesoscopic nonequilibrium stationary states, and describe a method to quantify irreversibility from the statistics of a single stochastic variable. In Sec.~\ref{sec:3}, we provide estimates of irreversibility from experimental measurements of hair-bundle fluctuations. In Sec.~\ref{sec:4} we compare the one-variable irreversibility estimates with  entropy production obtained from numerical simulations of a  stochastic model of active hair-bundle fluctuations. In Sec.~\ref{sec:5}, we use  thermodynamic uncertainty relations to predict how much entropy production can be  estimated by having access to the motors hidden state. Finally, in Sec.~\ref{sec:6} we  discuss our main findings and conclude the paper. Mathematical derivations, details on experimental data analysis and on biophysical modelling are provided in the Appendices.
 
 \section{Irreversibility and dissipation in stationary processes}
 \label{sec:2}
 
 \subsection{Generic properties}
We first discuss the relation between entropy production and irreversibility for generic nonequilibrium stationary processes. Consider a physical system described by a set of variables labeled as $X_{\alpha}$, with $\alpha= 1,2,\dots$. In a stationary nonequilibrium process of time duration $t$, the physical system traces a trajectory in the phase space described by the stochastic processes $X_{\alpha}(t)$. We denote by $\mathbf{\Gamma}_{[0,t]}\equiv \{ (x_{1}(s),x_{2}(s),\dots))\}_{s=0}^t$ a given trajectory described by the system variables and its corresponding time-reversed trajectory as  $\widetilde{\mathbf{\Gamma}}_{[0,t]}\equiv \{(\theta_1 x_{1}(t-s),\theta_2 x_{2}(t-s),\dots)\}_{s=0}^t$, where $\theta_\alpha=\pm 1$ is the time-reversal signature of the $\alpha-$th variable.  Assume now that $X_\alpha$ are the   variables  that may be out of equilibrium, i.e. we do not include in $\mathbf{\Gamma}_{[0,t]}$ those variables corresponding to thermal reservoirs, chemostats, etc. In that case, the steady-state rate of entropy production  $\sigma_{\textrm{tot}}$  is given by
\begin{equation}
\sigma_{\textrm{tot}} = k_{\rm B} \lim_{t\to\infty} \frac{1}{t}D\left[\mathcal{P}\Big(\mathbf{\Gamma}_{[0,t]}\Big)\right|\left|\mathcal{P}\left(\widetilde{\mathbf{\Gamma}}_{[0,t]}\right)\right]\quad,
\label{eq:1}
\end{equation}
where $\kb$ is the Boltzmann constant and $\mathcal{P}$ denotes the steady-state path  probability~\cite{lebowitz1999gallavotti,maes2003time,seifert2005entropy,neri2017statistics}. Here  $D[\mathcal{Q}||\mathcal{R}]\geq 0$ is the Kullback-Leibler (KL) divergence between the probability measures $\mathcal{Q}$ and $\mathcal{R}$, which quantifies the distinguishability between these two distributions. For measures of a single random variable $x$ the KL divergence is given by $D[\mathcal{Q}(x)||\mathcal{R}(x)]\equiv\int \text{d}x\, \mathcal{Q}(x)\ln [\mathcal{Q}(x)/\mathcal{R}(x)]$. Note that for isothermal systems $\sigma_{\rm tot}T$ equals to the rate of heat dissipated to the environment at temperature $T$.

Often in experiments only one or several of the nonequilibrium variables can be tracked in time. Consider the case where only 
$X_1,\dots X_k$ are known. We define the $k-$variable 
 irreversibility measure in terms of path probabilities of $k$ mesoscopic variables
\begin{equation}
\sigma_k \equiv k_{\rm B} \lim_{t\to\infty} \frac{1}{t}D\left[\mathcal{P}\left(\mathbf{\Gamma}^{(k)}_{[0,t]}\right)\right|\left|\mathcal{P}\left(\widetilde{\mathbf{\Gamma}}^{(k)}_{[0,t]}\right)\right]\quad, \label{eq:3}
\end{equation}
where $\mathbf{\Gamma}^{(k)}_{[0,t]}\equiv \{ (x_{1}(s),\dots,x_{k}(s))\}_{s=0}^t$ and $\widetilde{\mathbf{\Gamma}}^{(k)}_{[0,t]}\equiv \{( \theta_1x_{1}(t-s),\dots,\theta_k x_{k}(t-s))\}_{s=0}^t$ denote paths described by $k$ variables. The 
$k-$variable 
 irreversibility measure increases with the number of tracked degrees of freedom,  providing a set of lower bounds to entropy production:
 \begin{equation}
     0\leq  \sigma_1\leq\dots\leq \sigma_k  \leq \sigma_{k+1} \leq \dots\leq \sigma_{\rm tot}.
 \end{equation}
It can also be shown that the estimator $\sigma_k$ equals the physical entropy production $\sigma_{\rm tot}$ if the missing variables, $X_{\ell}$ with $\ell>k$, are at thermal equilibrium~\cite{gomez2008footprints,mehl2012role,celani2012anomalous}. When the missing variables are not at thermal equilibrium, which is often the case in active systems, the estimate $\sigma_k \leq \sigma_{\rm tot}$ yields only a lower bound for the  entropy production rate. 

\subsection{One variable irreversibility measure}

We now introduce a  method to estimate the irreversibility measure $\sigma_1$ for any nonequilibrium steady state from a single stationary time series $x_i=X(i\Delta t)$ ($i=1,\dots,n$) of a variable $X$ that is even under time reversal. We describe the technique  for a single variable, but it can be generalized to several variables $X_\alpha(t)$.
In discrete processes, the KL divergence in $\sigma_1$ can be accurately measured from the statistics of sequences of symbols~\cite{roldan2010estimating,roldan2012entropy}. In continuous processes however, estimating $\sigma_1$ is a herculean task due to the difficulties in sampling  the whole  phase space of paths~\cite{andrieux2008thermodynamic,tusch2014energy,roldan2014irreversibility}.

  The key idea of the method is to exploit the invariance of the KL divergence under  one-to-one transformations. Suppose that there exists a 
one-to-one 
map $\xi_i(x_1,\dots,x_n)$,  $i=1,\dots,n$, that transforms the original time series and its time reversal into two  new time series
$\xi^F_i=\xi_i(x_1,\dots,x_n)$ and $\xi^R_i=\xi_i(x_n,\dots,x_1)$ that are  independent and identically distributed (i.i.d.) processes. Such a procedure is often called a \textit{whitening} filter~\cite{efron1994detection,galka2006whitening}.
Because the new series are  i.i.d., the KL divergence is now simple to calculate: it is given by the KL divergence between two \textit{univariate} distributions $p(\xi)$ and $q(\xi)$, corresponding to the stationary probability distribution of $\xi_i^F$ and $\xi_i^R$, respectively~\cite{roldan2014irreversibility}. In general, it is not possible to find a one-to-one map that fully eliminates the correlations of both the forward $(x_1,\dots,x_n)$ and the backward $(x_n,\dots,x_1)$ time series. In that case, the removal of the correlations in the backward series is enough to provide a lower bound for $\sigma_1$:
\begin{equation}
\sigma_1 \geq k_{\rm B} f_{\rm s} D[p(\xi)||q(\xi)]\equiv \hat{\sigma}_1\quad,
\label{eq:finalbound}
\end{equation}
where $f_{\rm s}=(\Delta t)^{-1}$ is the sampling frequency and $D[p(\xi)||q(\xi)]=\int \text{d}\xi\, p(\xi)\ln[p(\xi)/q(\xi)]$ is the KL divergence between the  univariate distributions $p(\xi)$ and $q(\xi)$. We estimate $D[p(\xi)||q(\xi)]\simeq \gamma \sum_{i} \hat {p}_i \ln (\hat{p}_i/\hat{q}_i)$ where $\hat{p},\hat{q}$ are empirical densities,  and the sum runs  over the number of histogram bins. We introduce the prefactor $\gamma=1-p_{\rm KS}\leq 1$,  where $p_{\rm KS}$ is the p-value of the Kolmogorov-Smirnov statistic between  $p(\xi)$ and $q(\xi)$, to correct the statistical bias of our KL divergence estimate~\cite{bonachela2008entropy}.
The proof of the bound~\eqref{eq:finalbound} and further details of the estimate are found in~Appendices~\ref{app:i} and~\ref{app:ii}.


\section{One variable irreversibility in active hair-bundle fluctuations}
\label{sec:3}

\begin{figure}
\centering
\includegraphics[width=0.45\textwidth]{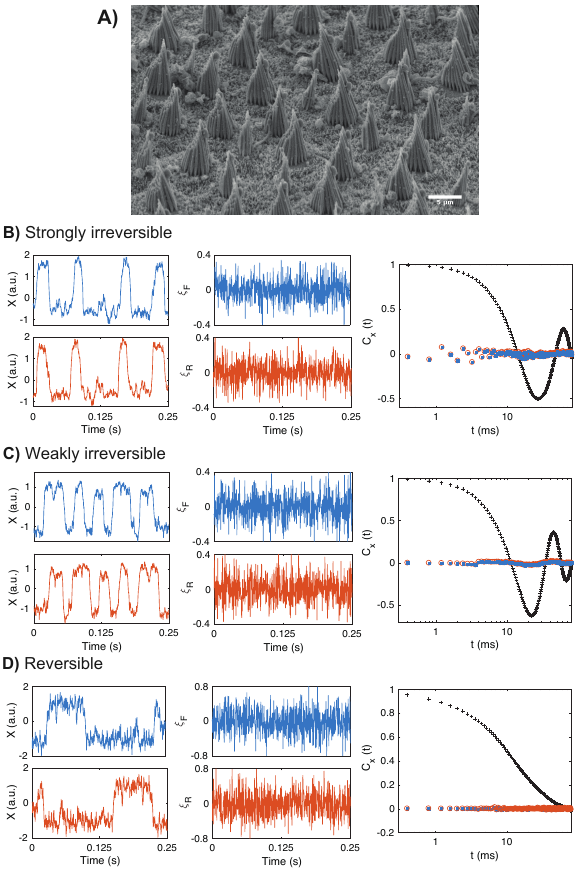}
\vspace{-0.2cm}
\caption{ \textbf{(A)} Electron micrograph showing hair bundles that protrude from the sensory epithelium of the bullfrog's sacculus. This  organ from the  frog's inner ear is dedicated to the detection of low frequency vibrations ($\sim 5-150$ Hz).  The height of the hair bundles is about $\sim7\mu\rm m$.  \textbf{(B-C)} Experimental recordings of the tip position  of two active mechanosensory hair bundles that display spontaneous oscillations. 
\textbf{(D)} Time series obtained from a simulation of a stochastic bistable oscillator, see text for details.  In panels \textbf{(B-D)}, we plot: the position $X$ of a hair bundle as a function of time (top left); their time reversals (bottom left);  residual time series  $\xi^F_i$ (top middle), $\xi^{R}_i$ (bottom middle); and the autocorrelation functions (right column) of the full $30$s recording (black "+"), $\xi^F_i$ (blue squares), and $\xi^{R}_i$ (red circles). The data in \textbf{(D)} corresponds to a stochastic simulation of a  system obeying the Langevin equation $\dot{x} = -V'(x) + \sqrt{2D}\xi$, with  the bistable potential  $V(x)=-ax^2/2 + bx^4/4$ and parameter values $a=30$, $b=0.4$, $D=20$~\cite{gammaitoni1998stochastic}.
 \label{fig:1}}
\end{figure}

We now discuss irreversibility and entropy production  in active mechanosensory hair cells from the bullfrog's ear. 
In experimental recordings of spontaneous hair-bundle oscillations, only the tip position $X_1$ of the  bundle is measured (Fig.~\ref{fig:1}B-C).   Hair-bundle oscillations take the shape of  relaxation oscillations corresponding: an alternation of fast jumps between two extreme positions interspaced by dwell times.
Measuring $X_1$, we can only estimate $\sigma_1$, which provides a lower bound to the total steady-state entropy production rate $\sigma_{\rm tot}$.   We later compare this estimate to that obtained for a passive bistable system in a thermal bath (Fig.~\ref{fig:1}D).

\begin{figure}
\centering
\includegraphics[width=0.35\textwidth]{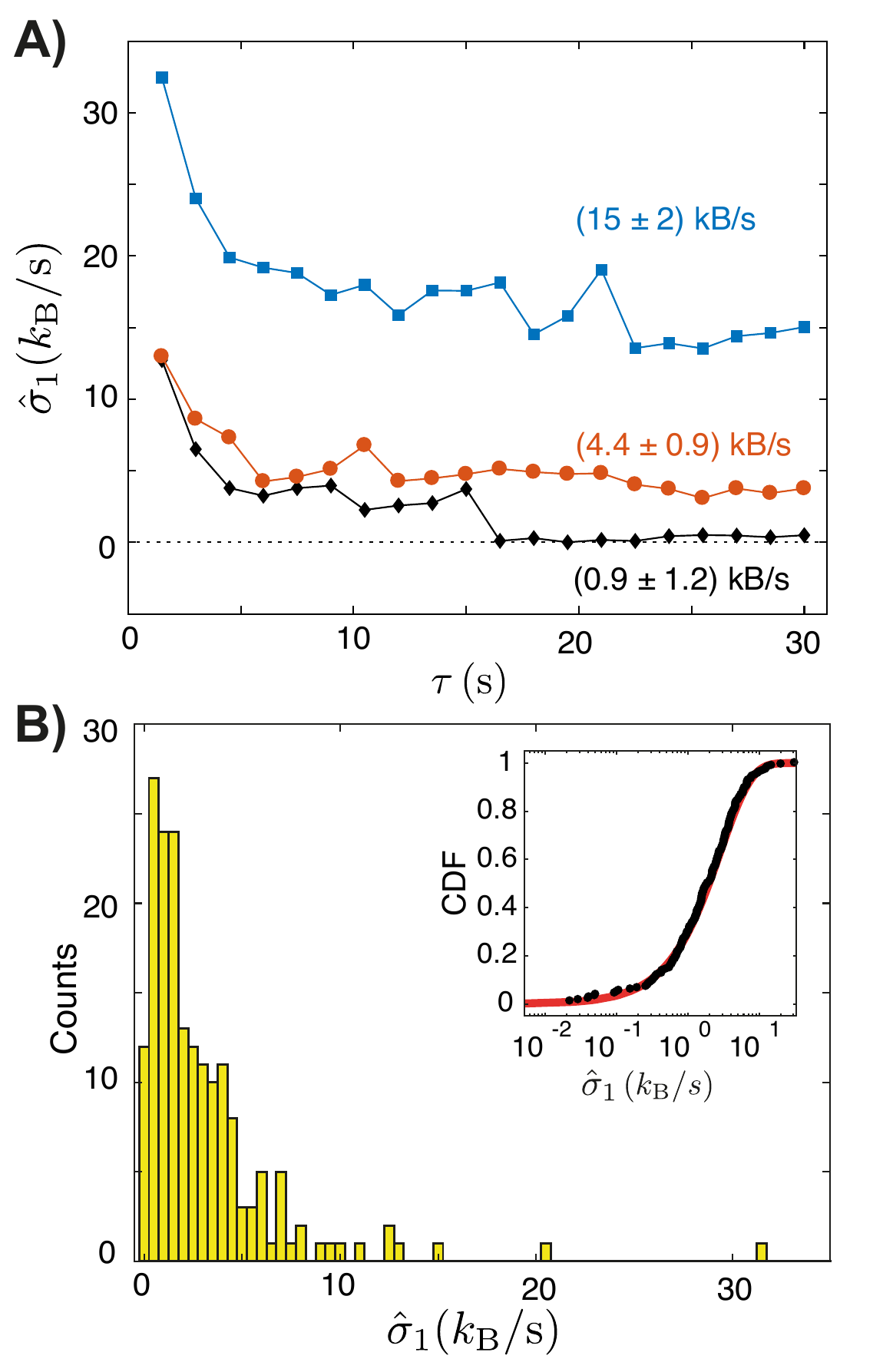}
\caption{  \textbf{(A)} Irreversibility  measure $\hat{\sigma}_1$ (symbols)  as a function of the observation time $\tau$ obtained from the time series partially shown in Fig.~\ref{fig:1}A. The horizontal dashed line is set to zero, corresponding to the reversible limit. The values of irreversibility obtained for these time series is indicated in the figure legends. They are given by the mean of the values of $\hat\sigma_1$ for $\tau > 10$s and the error bars  by the standard deviation of the same values. \textbf{(B)} Histogram of the irreversibility measure $\hat{\sigma}_1$ obtained from $182$ experimental recordings of spontaneous active oscillations of the hair bundle of duration $30\,\rm s$. The experimental average value of the irreversibility measure $\hat{\sigma}_1$ is $\sim 3k_{\rm B}/s$. Inset: Empirical cumulative distribution function (CDF) of irreversibility (black circles). The red line is a fit to  an exponential distribution 
 with mean value $(2.82 \pm 0.02)k_{\rm B}/s$ and $R^2 > 0.9990$. } \label{fig:2}
\end{figure}

In the following, we make use of autoregressive (AR) models for the whitening transformation. More precisely, we obtain the transformed time series $\xi_i^F$ ($\xi_i^R$) as the  difference between the observed values  of the forward (backward) time series and the forecast of that value based on an AR model of order $m=10$. Parameters of the model are determined from fits of the AR-model to the time-reversed series of positions. 
The residual time series $\xi_i^F$ and $\xi_i^R$ (Fig.~\ref{fig:1}B-D, center column)
obtained from the whitening transformation are uncorrelated (Fig.~\ref{fig:1}B-D, right column) and are therefore i.i.d. processes in good approximation. 

We find that the irreversibility measure $\hat{\sigma}_1$  given by Eq.~\eqref{eq:finalbound} distinguishes active hair-bundle fluctuations ($\hat{\sigma}_1>0$) from passive fluctuations of a bistable system  ($\hat{\sigma}_1\simeq 0$). Note that the estimate saturates to a plateau when the time series is long enough, in practice here longer than $10$s. Using a population of 182 hair cells that showed spontaneous hair-bundle oscillations~\cite{barral2018friction}, we obtain an exponential
 distribution of $\hat{\sigma}_1$  with mean value $3\,k_{\rm B}/\text{s}$ (Fig.~\ref{fig:2}B).
 Interestingly, this result depends on the sampling frequency $f_{\rm s}$ (see Appendix~\ref{app:iii}):  irreversibility is maximal in the range  $f_{\rm s}\sim (200-600)\text{Hz}$ where its value goes up to $4.3\,k_{\rm B}/\text{s}$. This frequency dependency may provide additional information about  timescales of the underlying active process~\cite{si}.

\begin{figure}
\centering
\includegraphics[width=0.4\textwidth]{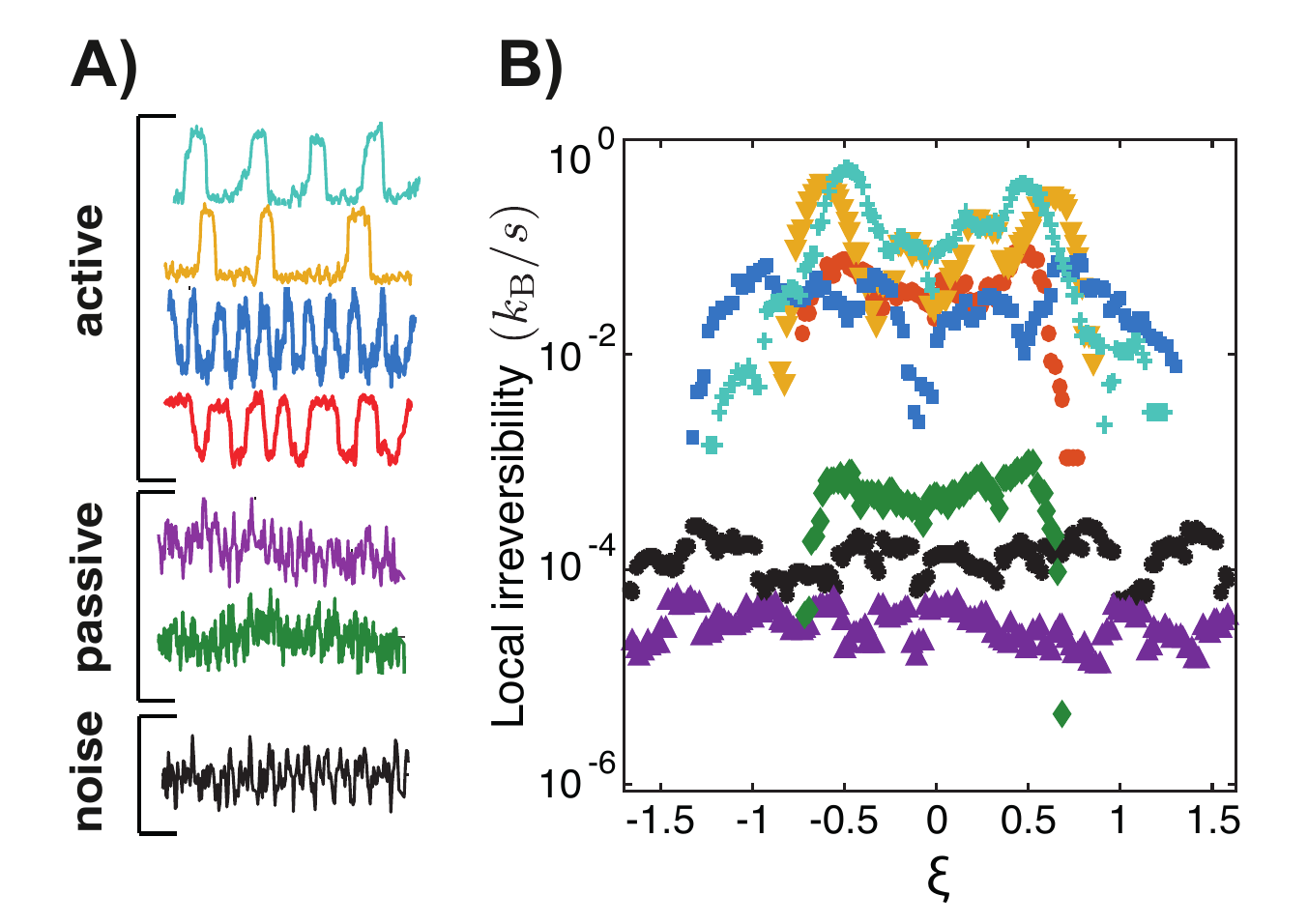}
\vspace{-0.2cm}
\caption{\textbf{(A)} Examples of experimental traces for the tip position   of different mechanosensory hair bundles as a function of time.  Top: active hair bundles. Bottom: passive hair bundles, i.e. when the channel blocker gentamicin is present (magenta, green), and experimental noise trace (black). 
\textbf{(B)} Estimate of the local irreversibility measure~\eqref{eq:lim} obtained from single  $30$s recordings of the oscillations shown in panel A as a function of the residual value $\xi$.
The sampling rate was $f_{\rm s}=2.5\,\text{kHz}$. 
 \label{fig:3}}
\end{figure}

We  further  quantify differences in irreversibility in typical examples of:  (i) active oscillatory hair bundles (Fig.~\ref{fig:3}A, top); (ii)  hair bundle that we were brought to  quiescence upon exposure to a drug (gentamicin) that blocks the transduction channels (Fig.~\ref{fig:3}A, middle); (iii)  noisy signals produced by the recording apparatus when there is no hair bundle under the objective of the microscope (Fig.~\ref{fig:3}A, bottom). 
To  further characterize  differences in irreversibility, we apply the local irreversibility measure defined as \begin{equation}
\hat{s}_1(\xi) \equiv k_{\rm B}f_{\rm s} \left[  p(\xi) \ln\frac{p(\xi)}{q(\xi)} + q(\xi)-p(\xi) \right]\quad,\label{eq:lim}
\end{equation}
which obeys $\hat{s}_1(\xi)\geq0$ for all $\xi$~\cite{shiraishi2016universal}, and $\hat{\sigma}_1=\int \text{d}\xi\hat{s}_1(\xi)$.  We find that for all the analyzed  values of $\xi$, the local irreversibility of active oscillations is $\sim\!10^{3}$ times larger than for passive oscillations and experimental noise.

\section{Entropy production rate of active hair-bundle fluctuations}
\label{sec:4}

\begin{figure}
\centering
\includegraphics[width=0.45\textwidth]{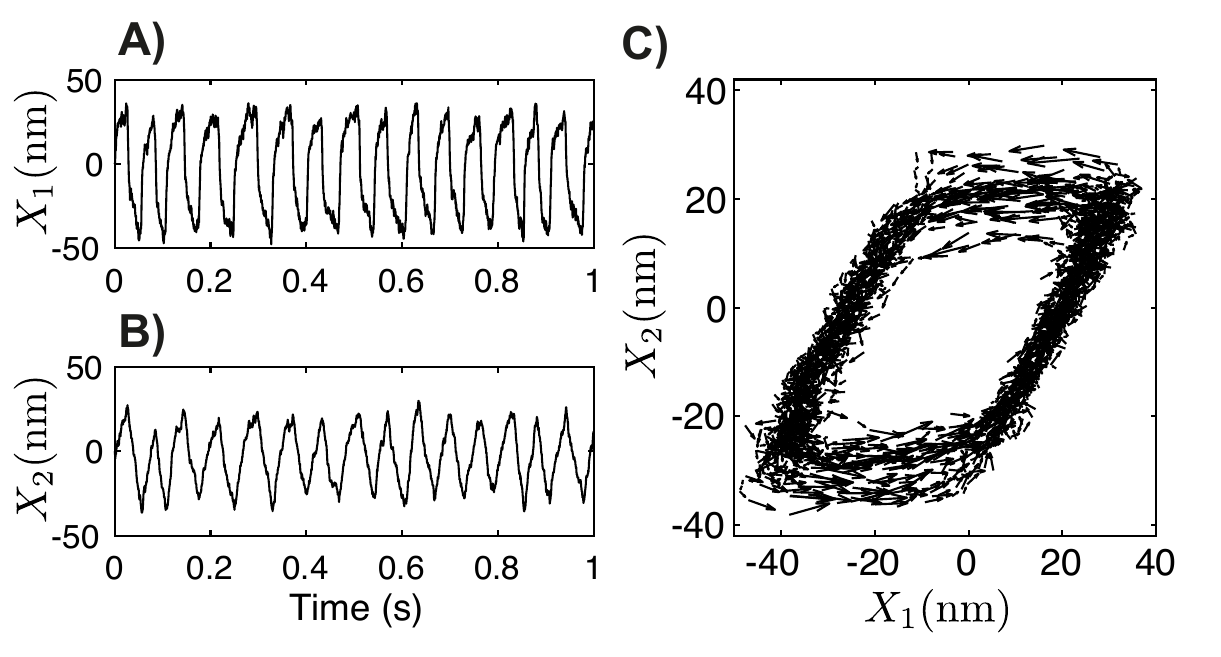}
\caption{   \textbf{(A,B)} Trajectories of the
reduced variables $X_1$ \textbf{(A)} and $X_2$ \textbf{(B)} as a function of time
obtained from a simulation of the stochastic model given by
Eqs.~(\ref{eq:4}-\ref{eq:5}). \textbf{(C)} Representation of a $2$-s trace of the simulations in \textbf{(A,B)} in the $\{X_1(t),X_2(t)\}$ plane. The black arrows illustrate the value of the instantaneous velocity and the base of the arrow the position. Parameters of the simulations: $\lambda_1=0.9 \,\rm pN ms/nm$, $\lambda_2=5 \,\rm pN ms/nm$, $k_{\rm gs} = 0.55 \, \rm pN/nm$, $k_{\rm sp} = 0.3\,\rm pN/nm$, $D=72\,\rm nm$, $S=0.73$, $F_{\rm max}=45.76\,\rm pN$, $N=50$,  $\Delta G=10 \kb T$, $\kb T=4.143\,\rm pN nm$ and $T_{\rm eff}/T=1.5$.
 \label{fig:1b}}
\end{figure}

We now relate the  estimate $\hat{\sigma}_1$ of entropy production from experimental recordings (Fig.~\ref{fig:2}B) to the entropy production $\sigma_{\rm tot}$ which we obtain from stochastic simulations of 
 hair-bundle oscillations.  Spontaneous hair-bundle oscillations are thought to result from an interplay between opening and closing of mechanosensitive ion channels, activity of molecular motors that pull on the channels, and fast calcium feedback. This interplay can be described by two coupled stochastic differential equations for the position of the bundle $X_1$ and of the  center of mass of a collection of molecular  motors $X_2$~\cite{tinevez2007unifying,bormuth2014transduction,barral2018friction} (see Appendix~\ref{app:vi}):
\begin{eqnarray}
\lambda_1\dot{X}_1 &=& -\frac{\partial V}{\partial X_1} + \sqrt{2k_{\rm B}T\lambda_1 }\;\xi_1 \label{eq:4}\\
\lambda_2 \dot{X}_2   &=& -\frac{\partial V}{\partial X_2} - F_{\rm act} + \sqrt{2k_{\rm B} T_{\rm eff}\lambda_2}\;\xi_2 \label{eq:5}\quad .
\end{eqnarray}
Here, $\lambda_1$ and $\lambda_2$ are friction coefficients and $\xi_1$ and $\xi_2$ in~(\ref{eq:4}-\ref{eq:5}) are two independent Gaussian white noises with zero mean $\langle \xi_i(t)\rangle =0$ ($i=1,2$) and correlation $\langle \xi_i(t)\xi_j(t')\rangle =\delta_{i j}\delta(t-t')$, with $i,j=1,2$ and $\delta_{ij}$ the Kronecker's delta. $T$ is the temperature of the environment, whereas the parameter $T_{\rm eff}>T$ is an effective temperature that characterizes fluctuations of the motors.
The conservative forces derive from the potential associated with elastic elements and mechano-sensitive ion channels
\begin{eqnarray}
V(X_1,X_2) &=&\frac{k_{\rm gs} \Delta X^2+k_{\rm sp}X_1^2}{2}  \\
&-& Nk_{\rm B}T\ln \left[ \exp\left(\frac{k_{\rm gs}D(X_1-X_2)}{Nk_{\rm B}T}\right)+A\right]\quad,\nonumber
\end{eqnarray}
where  $k_{\rm gs}$ and $k_{\rm sp}$ are stiffness coefficients; $D$ is the gating swing of a transduction channel; and $A=\exp [(\Delta G + (k_{\rm gs}D^2)/2N)/(k_{\rm B}T)]$, $\Delta G$ being the energy difference between open and closed states of the channels and $N$ the number of transduction elements.  The  force $F_{\rm act}(X_1,X_2) = F_{\rm max} (1-SP_{\rm o}(X_1,X_2))$ is an active nonconservative force exerted by the molecular motors with a maximum value $F_{\rm max}$. The parameter $S$ quantifies calcium-mediated feedback on the motor force~\cite{nadrowski2004active} and 
 \begin{equation}
 P_{\rm o}(X_1,X_2) =\frac{1}{1+A\exp(-k_{\rm gs}D(X_1-X_2)/Nk_{\rm B}T)}\quad,
 \label{eq:po}
 \end{equation}
  is the open probability of the transduction channels. 
    Note that Eq.~\eqref{eq:po} is the open probability of a  two-state equilibrium model of a channel that with a free energy difference between open and close states which depends linearly on the distance $X_1-X_2$. 
 As shown earlier~\cite{nadrowski2004active,tinevez2007unifying}, this model  can capture key features of noisy spontaneous oscillations of hair-bundle position $X_1$ that have been observed experimentally (Fig.~\ref{fig:1b}A).  The oscillation of the motors' position  (Fig.~\ref{fig:1b}B) is known in the model but hidden in experiments.  Trajectories of only $ X_1(t)$ or $X_2(t)$ do not reveal obvious signs of a net current, which here would correspond to a drift. However, trajectories in the $(X_1,X_2)$ plane show a net current which is a signature of entropy production (Fig.~\ref{fig:1}C). In the following, we will use this stochastic model to compare the irreversibility measure $\hat\sigma_1$ to the total entropy production $\sigma_{\rm tot}$.

 \begin{figure}
 \centering
\includegraphics[width=0.5\textwidth]{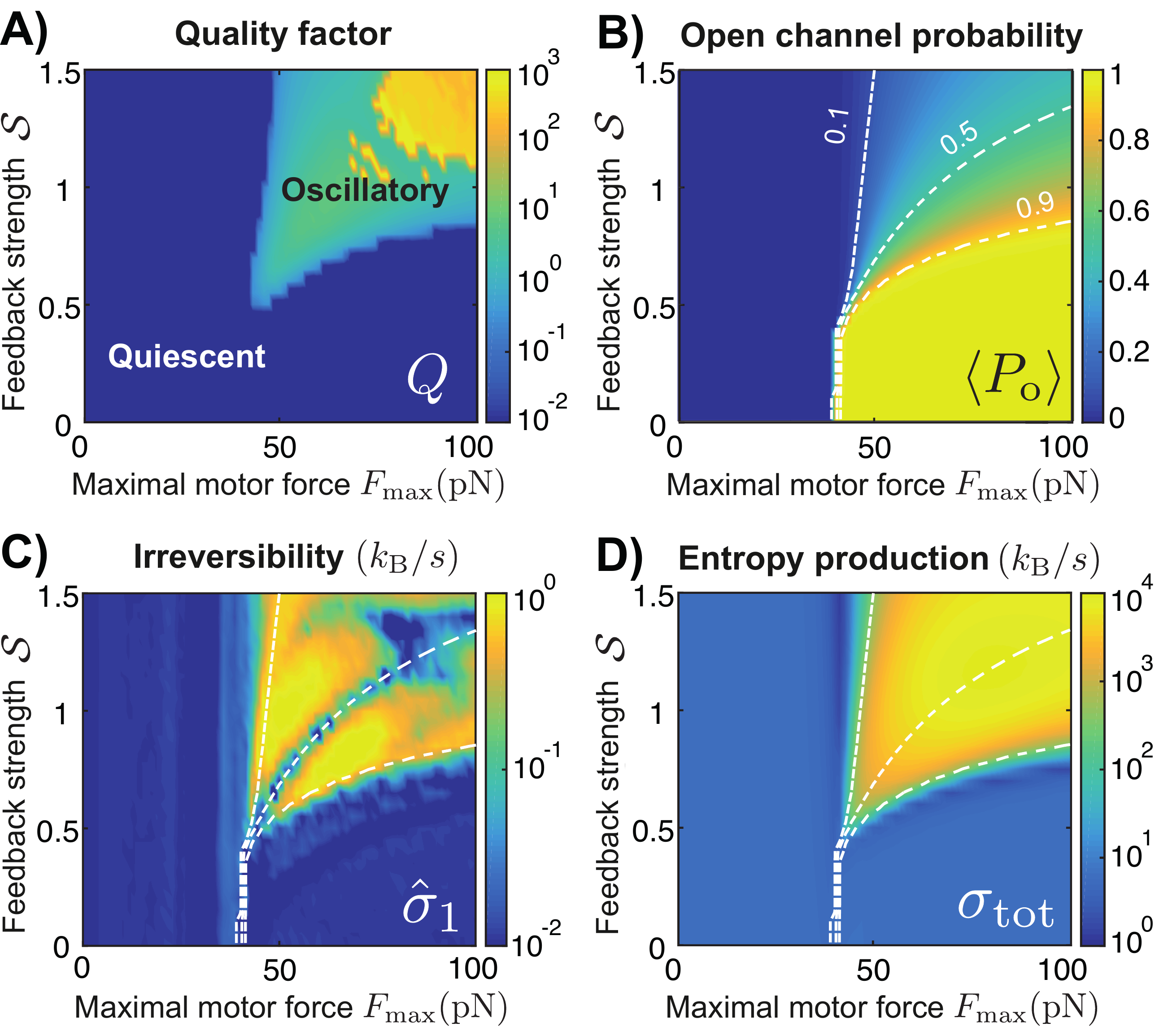}
\caption{Dynamical and thermodynamic features of spontaneous hair-bundle  oscillations as a function the calcium-feedback strength $S$ and maximal motor force $F_{\rm max}$ obtained from  numerical simulations of the model given by Eqs.~(\ref{eq:4}-\ref{eq:5}): \textbf{(A)} Quality factor $Q$; \textbf{(B)} Steady-state average of the open channel probability $\langle P_{\rm o}\rangle$; \textbf{(C)} Irreversibility measure $\hat{\sigma}_1$;  \textbf{(D)} Steady-state entropy production rate  $\sigma_{\rm tot}$. 
In (B,C,D) we indicate the parameter values for which $\langle P_{\rm o}\rangle = 0.1, 0.5$ and $0.9$ (white dashed lines from top to bottom, respectively). 
The  results  are obtained from numerical simulations of  Eqs.~(\ref{eq:4}-\ref{eq:5}) of total duration  $t_{\rm sim}=300\,\rm s$, sampling frequency $f_{\rm s}=1\,\rm kHz$ and parameter values $\lambda_1=2.8 \,\rm pN ms/nm$, $\lambda_2=10 \,\rm pN ms/nm$, $k_{\rm gs} = 0.75 \, \rm pN/nm$, $k_{\rm sp} = 0.6\,\rm pN/nm$, $D=61\,\rm nm$, 
$\Delta G=10 \kb T$, $\kb T=4 \,\rm pN nm$ and $T_{\rm eff}/T=1.5$. 
 \label{fig:4}}
\end{figure}
 In the stochastic model of hair-bundle oscillations given by Eqs. (\ref{eq:4}-\ref{eq:5}) we deal with only two variables, therefore $\sigma_{\rm tot}=\sigma_2$. 
 From the analytical expression of $\sigma_2$,  
 we find that the steady-state entropy production rate  can be written as~\cite{chetrite2008fluctuation,dabelow2018irreversibility} (see Appendix~\ref{app:iv})
\begin{equation}
\sigma_{\rm tot}= -\langle \dot{Q}_1\rangle \left(\frac{1}{T}-\frac{1}{T_{\rm eff}}\right) + \frac{\langle \dot{W}_{\rm act}\rangle}{T_{\rm eff}}\quad,
\label{eq:11}
\end{equation}
where $-\langle \dot{Q}_1\rangle  =- \langle (\partial V/\partial X_1)\,\circ\, \dot{X}_1\rangle$ is the steady-state average heat dissipated to the thermal bath at temperature $T$ and $\langle \dot{W}_{\rm act}\rangle  = -\langle F_{\rm act}\,\circ \,\dot{X}_2\rangle$ is the  power exerted by the active force on the motors. Here   $\langle\, \cdot\, \rangle $ denote  steady state averages and $\circ$  the Stratonovich product~\cite{sekimoto1998langevin,si}. Equation~\eqref{eq:11} reveals  two sources of nonequilibrium in the model: the difference of effective temperature and temperature, and the active force.

 \begin{figure*}
 \centering
\includegraphics[width=\textwidth]{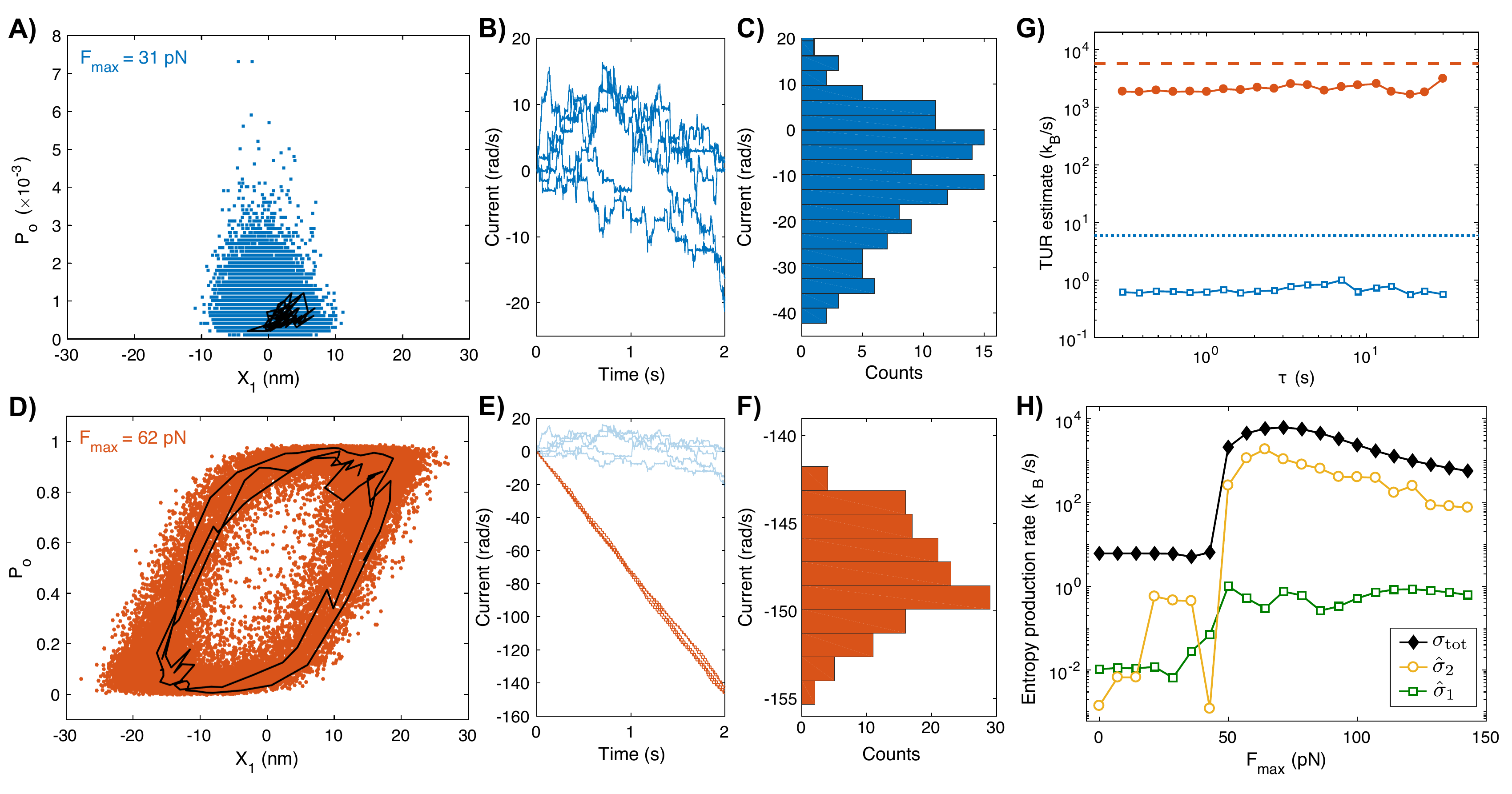}
\caption{Estimating entropy production using a thermodynamic uncertainty relation.     \textbf{(A,D)} Scatter plots of a sample trajectories $\mathbf{\Gamma}_{[0,\tau]}\equiv \{ (X_{1}(s),P_{\rm o}(s))\}_{s=0}^\tau$ of duration $\tau=30$s (symbols) obtained from a numerical simulations  of  Eqs.~(\ref{eq:4}-\ref{eq:5}). The black lines show the initial $0.3$s of the sampled trajectory. 
 \textbf{(B,E)} Sample trajectories of the counterclockwise current of the phase $\theta(t) = \tan^{-1}[\bar{P}_{\rm o}(t)/\bar{X}_1(t)] $  obtained from the reduced variables $\bar{P}_{\rm o}(t)= {P}_{\rm o}(t)-\langle{P}_{\rm o}(t)\rangle$ and   $\bar{X}_{1}(t)= {X}_{1}(t)-\langle{X}_{1}(t)\rangle$ for the quiescent (blue line) and oscillatory (red line) simulations. For comparison, we show in  \textbf{(E)} (light blue lines) the same trajectories displayed in \textbf{(B)}.   Note that,  even in the quiescent regime, we can detect a current in the $(X_1,P_{\rm o})$ space, revealing  activity in the fluctuations. These current fluctuations show  both smaller mean and larger relative uncertainty compared to those obtained with   simulations in the oscillatory  regime.
  \textbf{(C,F)}  Histograms of the cumulated current up to $\tau=2s$ obtained from in the quiescent (blue bars, \textbf{(C)}) and the  oscillatory (red bars, \textbf{(F)}) regimes.    \textbf{(G)} Comparison between the irreversibility estimate $\hat{\sigma}_2$ from the thermodynamic uncertainty relation (Eq.~\eqref{eq:tur}, symbols) and the total entropy production rate $\sigma_{\rm tot}$ (Eq.~\eqref{eq:11}, lines) as a function of the observation time $\tau$ in the quiescent (blue squares,  blue dotted line) and oscillatory (red circles, red dashed line)  regimes.  \textbf{(H)}  Comparison between the total entropy production rate $\sigma_{\rm tot}$ (black diamonds), the one-variable irreversibility measure $\hat{\sigma}_1$  (green squares), and the two-variable irreversibility measure $\hat{\sigma}_2$ from the thermodynamic uncertainty relation (Eq.~\eqref{eq:tur}, orange circles), as a function of the maximum motor force $F_{\rm max}$. In \textbf{(G,H)}  the lines are a guide to the eye. Simulation parameters: $\lambda_1=2.8 \,\rm pN ms/nm$, $\lambda_2=10 \,\rm pN ms/nm$, $k_{\rm gs} = 0.75 \, \rm pN/nm$, $k_{\rm sp} = 0.6\,\rm pN/nm$, $D=61\,\rm nm$, $\Delta G=10 \kb T$, $\kb T=4 \,\rm pN nm$, $T_{\rm eff}/T=1.5$, $S=0.94$, and simulation time step $\Delta t = 1$ms. Simulations were run for a total duration of   $300$s at the two operating points  with maximum motor force $F_{\rm max} = 31$pN \textbf{(A-C)} and $F_{\rm max} = 62$pN  \textbf{(D-F)}, corresponding to quiescent and oscillatory regimes, respectively.
  \label{fig:5}}
\end{figure*}

We performed numerical simulations of Eqs.~(\ref{eq:4}-\ref{eq:5}) for different values of the control parameters $F_{\rm max}$ and $S$ (Fig.~\ref{fig:1b}) to explore entropy production throughout  the state diagram of the system. The quality factor of the oscillation $Q$ $-$given by the ratio between the oscillation frequency and the bandwidth at half  the maximal height of the power spectrum (see Appendix~\ref{app:v})$-$ and the average open probability $\langle P_{\rm o} \rangle$ at steady state are displayed in Fig.~\ref{fig:4}A-B in the state diagram.
The irreversibility measure $\hat{\sigma}_1$ for  trajectories $X_1(t)$ of spontaneous oscillations is shown in Fig.~\ref{fig:1b}C. This measure can be compared to the quantification of total entropy production  $\sigma_{\rm tot}$ of the  model, given by Eq.~\eqref{eq:11}, which is shown in Fig.~\ref{fig:1b}D. Irreversibility of trajectories and total entropy production correlate strongly.  As expected, $\hat{\sigma}_1$ provides a lower bound to the actual dissipation rate.
Actually, the rate of entropy production estimated from $\hat{\sigma}_1$  is  here typically three orders of magnitude smaller than the total entropy production. Clearly, measuring other degrees of freedom additional to the hair-bundle position  would be required to obtain tighter bounds to the  rate of entropy production with our method or other estimation techniques~\cite{roldan2015decision,pietzonka2016universal,maes2017frenetic,li2019quantifying,frishman2018learning,van2020unified}.

\section{Thermodynamic uncertainty relation in the ear of the bullfrog}
\label{sec:5}

 Noisy limit-cycle oscillations in, for instance, a two-dimensional phase space can reveal irreversibility in the form of probability currents. It has been shown that the so-called thermodynamic uncertainty relations   provide   lower bounds to the rate of entropy production in terms of the mean and the variance of empirical time-integrated currents (see e.g.~\cite{barato2015thermodynamic,gingrich2016dissipation}).  Here, we  apply  one of these relations  to predict  how much entropy production one can assess by measuring two mesoscopic degrees of freedom:  the tip position $X_1$ of the hair bundle  and the  transduction  current,  normalized to its maximum value, $P_{\rm o}$ (see Eq.~\eqref{eq:po}).  Specifically, we analyze two-dimensional stochastic trajectories $\mathbf{\Gamma}_{[0,\tau]}\equiv \{ (X_{1}(s),P_{\rm o}(s))\}_{s=0}^\tau$ obtained from simulations of Eqs.~(\ref{eq:4}-\ref{eq:5}) in the quiescent (Fig.~\ref{fig:5}A) and oscillatory region (Fig.~\ref{fig:5}D) of the state diagram shown in Fig.~\ref{fig:4}.  These trajectories reveal a larger circulating probability current within the oscillatory region, as expected, but also a smaller relative uncertainty.

To quantify these effects, we map the dynamics into the complex plane $z(t)=\bar{X}_1(t) + \mathrm{i}\bar{P}_{\rm o}(t)$ and measure  $\theta(t)  = \phi(t) + 2\pi N_{\phi(t)}$, where $\phi(t)=\tan^{-1} (\bar{P}_{\rm o}(t)/\bar{X}_1(t))\in [0,2\pi]$ is the phase  and  $N_{\phi(t)}$ is the net number of counterclockwise turns---the winding number. Here,  $\bar{X}_{1}(t) = X_1(t)-\langle X_1\rangle$, $\bar{P}_{\rm o}(t)={P}_{\rm o}(t) -\langle P_{\rm o}\rangle$.  Using sample trajectories of duration $\tau=2$s, we found that the counterclockwise current $j(t)=\phi(t)/\tau$  displays both a larger absolute mean and a larger signal-to-noise ratio, corresponding to more accurate currents, when the system operates in the oscillatory (Figs.~\ref{fig:5}D) rather than in the quiescent regime of the dynamics (Figs.~\ref{fig:5}B).  Following Refs.~\cite{pietzonka2017finite,li2019quantifying}, the following thermodynamic uncertainty relation holds for any observation time window $\tau$:
\begin{equation}
\sigma_{\rm tot}  \frac{ \text{Var}[ j(\tau) ]}{ \langle j(\tau)\rangle^2 } \tau  \geq  2 k_{\rm B} \quad,
\label{eq:tur1}
\end{equation}
where $\text{Var}[ j(\tau) ]=\langle j^2(\tau)\rangle- \langle j(\tau)\rangle^2$ is the finite-time variance of the current. From Eq.~\eqref{eq:tur1},  we extract the estimate (see also Ref.~\cite{li2019quantifying,seara2021irreversibility})
\begin{equation}
\hat{\sigma}_2 \equiv \frac{2 k_{\rm B}}{\tau}\frac{ \langle j(\tau)\rangle^2 }{ \text{Var}[ j(\tau) ]} \quad,
\label{eq:tur}
\end{equation}
 which provides a lower bound for the total entropy production $\hat{\sigma}_2\leq \sigma_{\rm tot}$.  Note that the inequality~\eqref{eq:tur1} holds for any time-integrated current of a Markovian nonequilibrium steady state, which includes the one we measure as a particular case. 
We show estimates $\hat{\sigma}_2$  for the two case studies in (Fig.~\ref{fig:5}G).  For an example trajectory in the quiescent regime of the dynamics, $\hat{\sigma}_2\sim 1k_{\rm B}/$s is of the same order of magnitude as $\hat{\sigma}_1$ (Fig.~\ref{fig:5}G, blue squares).  Remarkably, operating in the oscillatory regime instead yields an estimate $\hat{\sigma}_2\sim 10^3 k_{\rm B}/$s (Fig.~\ref{fig:5}G, red circles), which is three orders of magnitude larger than $\hat{\sigma}_1$ and only a few fold smaller than $\sigma_{\rm tot}$. 

To get further insights on  entropy production upon varying the operating point in the state diagram of the system, we plot $\hat{\sigma}_2$ as a function of the maximal motor force $F_{\rm max}$ at fixed $S=0.94$ (Fig.~\ref{fig:5}H).  In the quiescent region, $\hat{\sigma}_2$ is not significantly different from $\hat{\sigma}_1$, predicting low entropy production ($\sim1k_{\rm B}/s$) about one order of magnitude below $\sigma_{\rm tot}$. Increasing $F_{\rm max}$, the two-variable irreversibility measure $\hat{\sigma}_2$ and the total entropy production $\sigma_{\rm tot}$ both exhibit a  jump when the system enters the oscillatory region of the dynamics, which is indicative of the underlying deterministic Hopf bifurcation, as also observed for other oscillatory systems in Ref.~\cite{seara2021irreversibility}. The one-variable irreversibility measure $\hat{\sigma}_1$ also increased in this region but the variation was smoother. 


\section{discussion}
\label{sec:6}
In this work, we have have shown that fluctuations of active systems can reveal the arrow of time even in the absence of net drifts or currents. The hierarchy of measures of time irreversibility introduced here provides lower bounds for the entropy production of an active process. We have demonstrated the applicability of the approach by estimating entropy production associated with experimental noisy oscillations of a single degree of freedom in the case of mechanosensory hair bundles from the bullfrog's ear. We have shown that quantifications of the arrow of time can efficiently discriminate quiescent and oscillatory hair bundles, as well as reaveal transitions between the two regimes in response to changes in a control parameter (e.g. Calcium concentration as in Ref.~\cite{tinevez2007unifying}). However, using a model of active hair bundle oscillations, we also showed that estimating the rate of entropy production with only one degree of freedom yields a lower bound that can be orders of magnitude smaller than the total entropy production rate in the system. In the case of hair-bundle oscillations, we predict that measuring a second degree of freedom, e.g. the transduction current, would add sufficient information to get a tight bound. With two degrees of freedoms, the current in the phase space and its fluctuations can be used to bound entropy production by means of thermodynamic uncertainty relations.  Overall, our results show that  irreversibility measures can quantify  entropy production in active matter, including living systems, from fluctuations of only a few mesoscopic degrees of freedom.

\begin{acknowledgments}
The electron micrograph of the hair bundles shown in  Fig.~\ref{fig:1}A was obtained by Atitheb Chaiyasitdhi, a PhD student in P Martin's group. We acknowledge stimulating discussions with Roman Belousov, Izaak Neri, Andre C. Barato, Simone Pigolotti,  Johannes Baumgart, Jose Negrete Jr,  Ken Sekimoto, Ignacio A. Mart\'inez,  Patrick Pietzonka and A.J. Hudspeth.\looseness-1 
\end{acknowledgments}
\vspace{-0.5cm}
 


\onecolumngrid
\newpage 

\appendix 

\section*{APPENDIX}

Here we present additional details of the methods and results discussed in the Main Text.
In Secs.~\ref{app:i} and \ref{app:ii}, we provide a derivation of the bound used in Eq.~\eqref{eq:finalbound} in the Main Text, and describe the whitening transformation that we use to estimate irreversibility of stochastic time traces.  
In Sec.~\ref{app:iii}, we analyze how our irreversibility measure depends on the data sampling rate of the experimental recordings of hair-bundle spontaneous fluctuations.
In Sec.~\ref{app:vi}, we discuss the biophysical model of hair-bundle oscillations and  the experimental techniques.
In Sec.~\ref{app:iv}, we discuss how entropy production is estimated in numerical simulations of the hair-bundle biophysical model. Section~\ref{app:v} provides details on the calculation of the quality factor of spontaneous oscillations shown in Fig.~\ref{fig:4}A in the Main Text.

\section{Bounds on the multivariate Kullback-Leibler divergence}\label{KLDbounds}
\label{app:i}

Here we prove a general lower bound for the Kullback-Leibler (KL) divergence between two multivariate probability densities $P_X(x_1,\dots,x_n)$ and  $Q_X(x_1,\dots,x_n)$ that fulfill the following: there exits a one-to-one map $\xi_i=\xi_i(x_1,\dots,x_n)$  with $i=1,\dots,n$, such that
\begin{enumerate}
\item  the transformed variables $\xi_i$ are identically distributed under both $P$ and $Q$, that is, the distributions $P_\Xi(\xi_1,\dots,\xi_n)$ and $Q_\Xi(\xi_1,\dots,\xi_n)$ have, respectively, identical marginal distributions $p(\xi)$ and $q(\xi)$ for any $\xi_i$ ($i=1,\dots, n$);
\item the transformed variables $\xi_i$  are independent and identically distributed (i.i.d.) under the distribution $Q$, that is, $Q_\Xi(\xi_1,\dots,\xi_n)=\Pi_i\, q(\xi)$.
\end{enumerate}

\noindent
The first step in the derivation is a simple application of the invariance of the KL distance under a one-to-one map:
\begin{equation}\label{eq:sini}
D\left[P_X(x_1,\dots,x_n) || Q_X(x_1,\dots,x_n)\right] =D\left[P_\Xi(\xi_1,\dots,\xi_n) || Q_\Xi(\xi_1,\dots,\xi_n)\right]\quad.
\end{equation}
Second, we can rewrite the relative entropy as
\begin{align}
 D\left[P_\Xi(\xi_1,\dots,\xi_n)||Q_\Xi(\xi_1,\dots,\xi_n)\right]  &= \int {\rm d}\xi_1\dots \int {\rm d}\xi_n \,
 P_\Xi(\xi_1,\dots,\xi_n)\ln\frac{P_\Xi(\xi_1,\dots,\xi_n)}{\Pi_i\,  q(\xi_i) }\nonumber
\\
 & =
 \int {\rm d}\xi_1\dots \int {\rm d}\xi_n \, 
\left[ 
 P_\Xi(\xi_1,\dots,\xi_n)\ln\frac{\Pi_i\,  p(\xi_i)}{\Pi_i\,  q(\xi_i) }+
 P_\Xi(\xi_1,\dots,\xi_n)\ln\frac{P_\Xi(\xi_1,\dots,\xi_n)}{\Pi_i\,  p(\xi_i) }\right]\nonumber
\\
&= nD[p(\xi)||q(\xi)]
+D\left[P_\Xi(\xi_1,\dots,\xi_n)|| \Pi_{i}\,  p(\xi_i)\right]\quad.\label{eq:s13a0} 
\end{align}
Because the KL divergence  between two distributions is always positive, Eqs.~\eqref{eq:sini} and~\eqref{eq:s13a0} yield the bound \begin{equation}
D\left[P_X(x_1,\dots,x_n) || Q_X(x_1,\dots,x_n)\right] \geq  nD[p(\xi)||q(\xi)]\quad,
\label{eq:s13a}
\end{equation}
and the inequality saturates if the transformed variables $\xi_i$ ($i=1\dots n$) are also i.i.d.~under $P_\Xi(\xi_1,\dots,\xi_n)$, i.e. when $ P_\Xi(\xi_1,\dots,\xi_n)=\Pi_i \, p(\xi_i)$. If one can find a one-to-one map that transforms the original random variables into i.i.d. variables under {\em both} distributions $P$ and $Q$, then \eqref{eq:s13a} becomes an equality and the exact KL divergence between the two multivariate distributions $P_X$ and $Q_X$ can be reduced to the KL divergence between single variable distributions $p(\xi)$ and $q(\xi)$, which is much easier to evaluate from real data. This is the key idea of our method to estimate the irreversibility of experimental time series.

\section{Irreversibility in continuous time series: the whitening transformation}
\label{app:ii}

The estimation of the KL divergence rate from single stationary trajectories of both discrete and continuous random variables have been previously discussed~\cite{roldan2014irreversibility}. For continuous random variables, the most common strategy is to make a symbolization or discretization of the time series~\cite{andrieux2008thermodynamic}. Then, the KL divergence is estimated from the statistics of substrings of increasing length~\cite{roldan2010estimating,roldan2012entropy}. The main limitation of this method is that one easily reaches lack of statistics even for short substrings. If the observed time series is non-Markovian, this limitation could yield inaccurate bounds for the entropy production. For instance, the KL divergence between two data substrings can be zero in non-equilibrium stationary states without observable currents \cite{roldan2014irreversibility,roldan2010estimating,roldan2012entropy}.

Here we introduce a new method to estimate the KL divergence rate
\begin{equation}
\frac{\sigma_1 }{k_{\rm B} }  \equiv \lim_{t\to\infty} \frac{1}{t}D\left[\mathcal{P}\left(\{x(s)\}_{s=0}^t\right)||\mathcal{P}\left(\{x(t-s)\}_{s=0}^t\right)\right]\quad,
\label{eq:sigma1}
\end{equation}
that is valid for continuous and possibly non-Markovian stochastic processes $X(t)$. First, in practice one has access to discrete-time  observations of the process $x_i \equiv X(i\Delta t)$, $i=1,\dots, n$, i.e., a time series  containing $n=t/\Delta t$ consecutive samples of the process with sampling rate $f_{\rm s}=1/\Delta t$. The time discretization implies a loss of information yielding a lower bound to the KL divergence rate:
\begin{equation}
\frac{\sigma_1 }{k_{\rm B} }  \geq f_{\rm s} \lim_{n\to\infty} \frac{1}{n}D[{P}_X(x_1,\dots, x_n)||Q_X(x_1,\dots,x_n)]\quad,
\label{eq:sigma1b}
\end{equation}
where $Q_X(x_1,\dots,x_n)=P_X(x_n,\dots,x_1)$ is the probability to observe the reverse trajectory $(x_n,\dots,x_1)$.

We can now apply the inequality \eqref{eq:s13a} to the right-hand side in Eq.~\eqref{eq:sigma1b} 
To do that, it is necessary to find a one-to-one map $\xi_i=\xi_i(x_1,\dots,x_n)$ that transforms the reverse time series $(x_n,\dots,x_1)$ into a sequence of $n$ i.i.d.~random variables, that is, into a white noise. Such a transformation  is usually termed {\em whitening transformation}. 

An example of whitening transformation is the time series formed by the residuals of an  autoregressive model, which is the transformation that we will use along this paper. A discrete-time stochastic process ${Y}_i$ is called  autoregressive of order $m$, AR$(m)$, when its value at a given time is given by a linear combination of its $m$ previous values plus a noise term. Such process is univocally determined by $m\geq 1$ real coefficients, $a_1, a_2, \dots, a_m$, a discrete-time white noise $\eta_i$ and a set of initial values $Y_1,Y_2,\dots, Y_m$. The values of $Y_i$ for $i>m$ are given by the linear recursion
\begin{equation}\label{eq:ARmodel}
{Y}_{i} = \sum_{j=1}^m a_j {Y}_{i-j} + \eta_i\quad. 
\end{equation}
Inspired by the AR$(m)$ process, we introduce the following one-to-one map
\begin{equation}
\xi_{i} =\begin{cases} x_i & \mbox{if $i\leq m$} \\ \displaystyle
 x_i - \sum_{j=1}^m a_j x_{i-j} & \mbox{if $i> m$}\end{cases}\quad,
\label{eq:residualseries}
\end{equation}
which is a linear transformation defined by a unitriangular matrix with Jacobian equal to one. With an appropriate choice of the coefficients $a_j$, one can get a new process $(\xi_1,\dots,\xi_n)$ which is approximately i.i.d. A good choice is given by a maximum likelihood fit of the process to the AR$(m)$ model. In that case, the elements $\xi_{i}$ in this new time series for $i>m$ are usually called {\em residuals} of the original time series $(x_1,\dots,x_n)$ with respect to the AR($m$) model. Notice also that, if $(x_1,\dots,x_n)$ is indeed a realization of the stochastic process~\eqref{eq:ARmodel}, then the residuals  are i.i.d. random variables and the process $(\xi_{m+1},\dots,\xi_n)$ has correlations $\langle \xi_{i} \xi_{j}\rangle = \delta_{ij}$ for all $i,j>m$.

 We now apply the bound \eqref{eq:s13a} to the KL divergence in the right hand side of Eq.~\eqref{eq:sigma1b}, using the transformation defined by Eq.~\eqref{eq:residualseries}. Since the contribution of the first, possibly correlated, $m$ values of the time series $\xi_i$, vanishes in the limit $n\to\infty$, we obtain the following lower bound to the KL divergence rate [Eq. (7) in the Main Text]: 
\begin{equation}
\frac{\sigma_1 }{k_{\rm B} }  \geq f_{\rm s} D[p(\xi)||q(\xi)]\quad.
\label{eq:sifinalbound}
\end{equation}

We can obtain empirical estimates of $p(\xi)$ and $q(\xi)$ from a single stationary time series $(x_1,\dots,x_n)$ as follows. We apply the transformation \eqref{eq:residualseries} to {\em both} the original time series $(x_1,\dots,x_n)$ and to its time reversal $(x_n,\dots,x_1)$ obtaining, respectively, two new  time series $(\xi^{\rm F}_1,\dots,\xi^{\rm F}_n)$ and $(\xi^{\rm R}_1,\dots,\xi^{\rm R}_n)$, which are stationary at least for $i>m$. The empirical PDFs obtained from the data of each series are estimations of, respectively, $p(\xi)$ and $q(\xi)$.
Note that the same transformation \eqref{eq:residualseries} must be applied to both  the original time series $(x_1,\dots,x_n)$ and its time reverse $(x_n,\dots,x_1)$, but the inequality \eqref{eq:s13a} only requires uncorrelated residuals \textit{in  the reverse series}. For this purpose, we calculate the coefficients $a_1,\dots,a_m$  by fitting the reverse time series $(x_n,\dots,x_1)$ to the AR$(m)$ model in Eq.~\eqref{eq:ARmodel}. 

As indicated in the previous section, the inequality \eqref{eq:sifinalbound}  is tighter when the residuals are uncorrelated in the forward series as well. This is the case of the experimental series that we have analyzed (see, for instance, Fig.~2B in the Main Text) although, in principle,  it is not guaranteed by this procedure. We remark that the inequality  \eqref{eq:sifinalbound} is a rigorous result if the transformation \eqref{eq:residualseries} applied to the reverse time series yields an uncorrelated series  $(\xi^{\rm R}_1,\dots,\xi^{\rm R}_n)$. 
In that case, $k_{\rm B} f_{\rm s} D[p(\xi)||q(\xi)]$ is an estimate of $\sigma_1$ with only two possible sources of error: {\em i)} the discrete sampling of the process $X(t)$ and  {\em ii)} the remnant correlation time in the  residuals $(\xi^{\rm F}_1,\dots,\xi^{\rm F}_n)$ obtained from the forward time series.
\newline\newline
To summarize, our theory provides an estimate $\hat{\sigma}_1$ for the KL divergence rate $\sigma_1$ which can be evaluated as follows: 
\begin{enumerate}
\item  Estimate the  coefficients, $a_1,\dots,a_m$, by fitting the \textit{time-reversed} series $(x_n,\dots,x_1)$ to an autoregressive AR($m$) model of order $m>1$.  A reasonable choice is $m=10$, but it should be tuned to minimize the correlation time in the residuals $(\xi_1^{\rm R},\dots,\xi_n^{\rm R})$.
\item  Apply the whitening transformation \eqref{eq:residualseries}  to the original series  $(x_1,\dots,x_n)$  and to its time reversal $(x_n,\dots,x_1)$ to obtain, respectively, new time series $(\xi_1^{\rm F},\dots,\xi_n^{\rm F})$ and $(\xi_1^{\rm R},\dots,\xi_n^{\rm R})$. 
Note that the new processes are not each other's time reversal.
\item  Obtain the empirical distributions $p(\xi)$  and $q(\xi)$ from the time series $(\xi_1^{\rm F},\dots,\xi_n^{\rm F})$ and $(\xi_1^{\rm R},\dots,\xi_n^{\rm R})$, respectively.
\item Calculate the KL divergence between $p(\xi)$ and   $q(\xi)$ 
\begin{equation}
D[p(\xi)||q(\xi)] = \int \text{d}\xi\, p(\xi)\ln \frac{p(\xi)}{q(\xi)}\quad,\label{eq:s9}
\end{equation}
 which can be estimated from numerical integration of the right hand side in~\eqref{eq:s9} using the empirical normalized histograms $p(\xi)$ and $q(\xi)$.  We call this estimate $\hat{\mathcal{D}}$, which is given by
 \begin{equation}
 \hat {\mathcal{D}}  = \gamma\sum_{i} \hat{p}_i  \ln \frac{ \hat{p}_i }{ \hat{q}_i }\quad, \label{eq:s102}
 \end{equation}
 where $\hat{p}_i=n_i^{\rm F}/(\sum_{i} n_i^{\rm F})$  and  $\hat{q}_i=n_i^{\rm R}/(\sum_{i} n_i^{\rm R})$ are the empirical probabilities,  obtained from 
the number of times  $n_i^{\rm F}$ and  $n_i^{\rm R}$  that the sequences $(\xi_1^{\rm F},\dots,\xi_n^{\rm F})$ and $(\xi_1^{\rm R},\dots,\xi_n^{\rm R})$ lie in the $i-$th bin, respectively. The sum in~\eqref{eq:s102} runs over all bins for which $n_i^{\rm F} >0$ and $n_i^{\rm R} >0$. For simplicity, we used $100$ bins of equal spacing ranging from the minimum to the maximum values of the residual time series $(\xi_1^{\rm F},\dots,\xi_n^{\rm F})$.

The value of the estimate  $\hat{\mathcal{D}}$ of the KL divergence~\eqref{eq:s9} is  weighted by a prefactor $\gamma\leq 1$ defined in terms of the probability to reject the null hypothesis $p(\xi)= q(\xi) $.  We use this procedure to correct  the statistical bias in the estimation of the KL divergence that appears when two stochastic processes have similar statistics~\cite{bonachela2008entropy,roldan2012entropy}. For this purpose, we use the Kolmogorov--Smirnov (KS) statistical test under the null hypothesis  $H_0: p(\xi) = q(\xi)$ which yields a   p-value $p_{\rm KS}$ for the two distributions to be equal. Here, small $p_{\rm KS}$  means that there is stronger statistical evidence in favour of the alternative hypothesis $p(\xi) \neq  q(\xi)$,   thus $\gamma = 1-p_{\rm KS}$ serves as a weight of irreversibility: $\gamma\simeq 0$ when it is hard to reject $H_0$ (reversibility) and $\gamma\simeq 1$ there is a larger statistical evidence to reject $H_0$.  
\item Finally, our estimate of $\hat{\sigma}_1$ is thus given by the KL divergence estimate $\hat{\mathcal{D}}$ times the Boltzmann constant and the data sampling frequency:
 \begin{equation}
 \hat{\sigma_1} = k_{\rm B} f_{\rm s}  \hat{\mathcal{D}}\quad. \label{eq:estimate}
 \end{equation}
\end{enumerate}

\section{Dependency of the irreversibility measure on the  sampling frequency}
\label{app:iii}
\begin{figure*}[h!]
\centering
\includegraphics[width=0.85\textwidth]{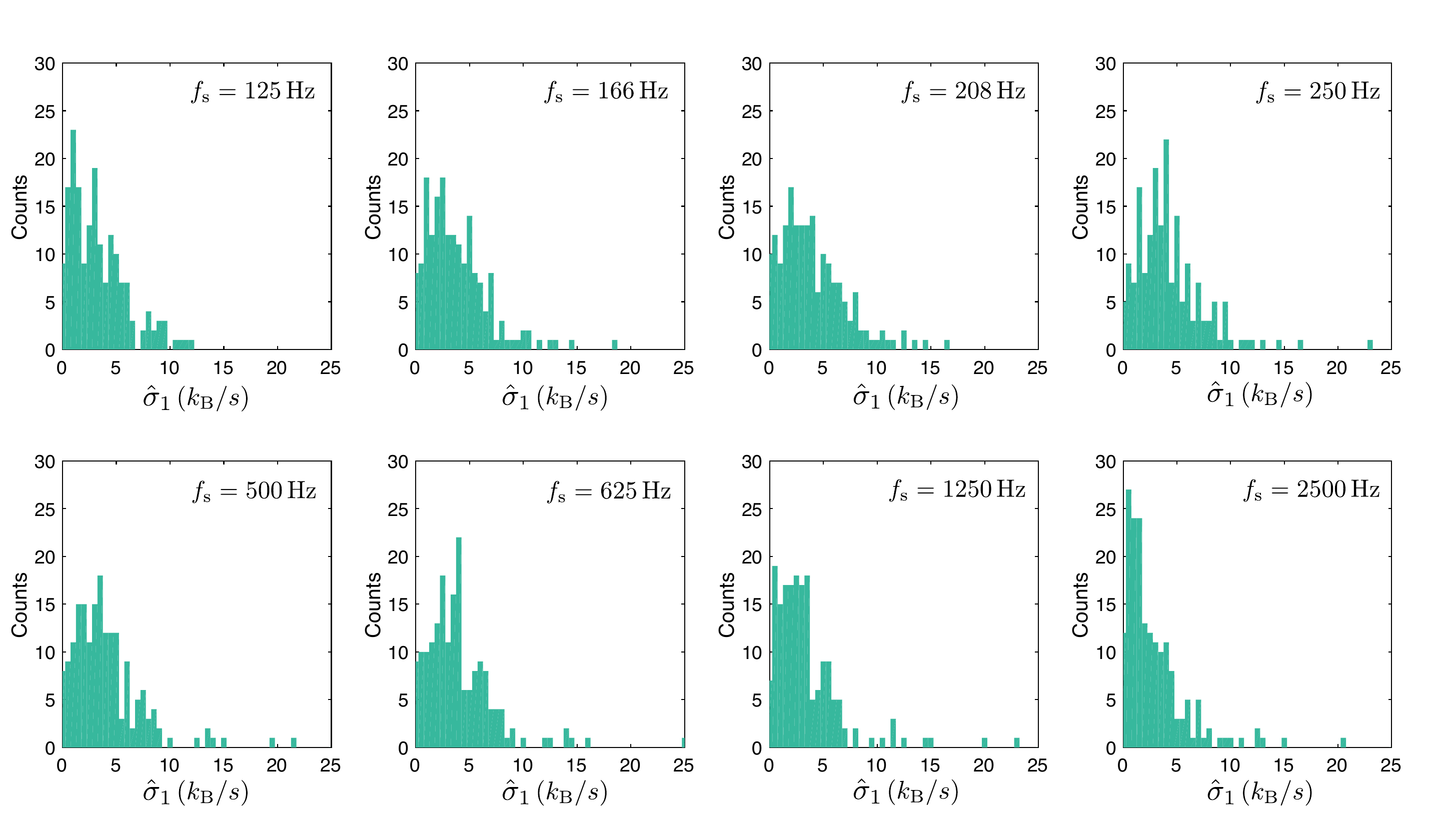}
\caption{Histograms of the irreversibility measure $\hat{\sigma}_1/k_{\rm B}$ for different values of the sampling frequency $f_{\rm s}$  indicated above each panel of the figure. All the histograms were obtained from the same ensemble of $182$ oscillatory hair cells that displayed active oscillations.  }
\label{fig:hists}
\end{figure*}

In this section, we analyse the dependency of our irreversibility measure on the sampling frequency $f_{\rm s}$.  For this purpose, we evaluate $\hat{\sigma}_1$ defined in Eq.~\eqref{eq:estimate} for $30$s recordings of the $182$ cells that showed spontaneous oscillations at different sampling frequencies, ranging from $125$Hz to $2500$Hz (the latter corresponding to  the data shown in Fig.~3C in the Main text). Figure~\ref{fig:hists} shows that the distribution of the irreversibility measure depends strongly on the sampling frequency of the data. 
Notably, the distributions shift towards higher irreversibility when the sampling frequency is reduced, until there is too much filtering $f_{\rm s}<250$Hz such that oscillations cannot be distinguished clearly. 

\begin{figure*}[h!]
\centering
\includegraphics[width=0.85\textwidth]{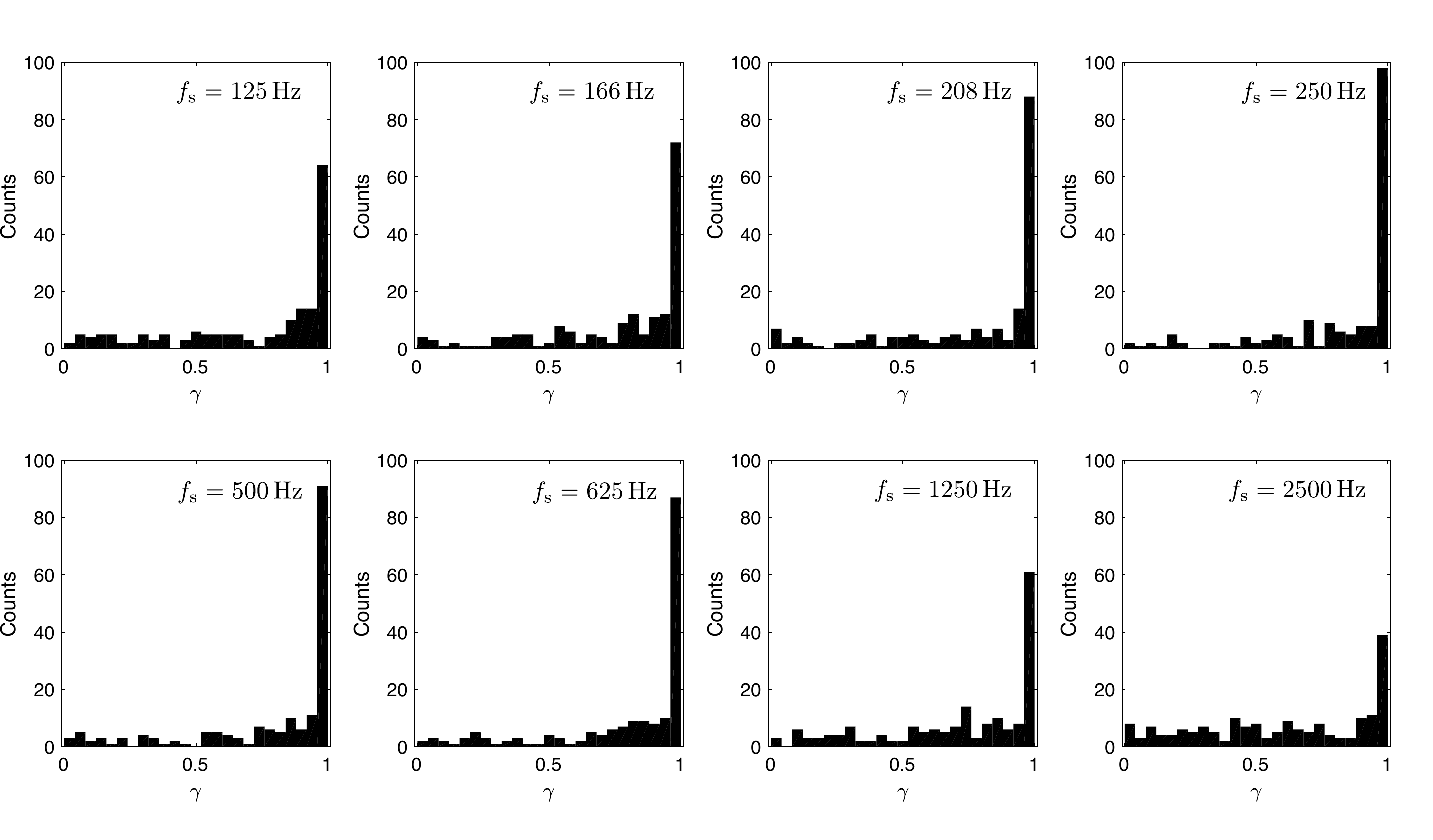}
\caption{Histograms of the parameter $\gamma=1-p_{\rm KS}$ with $p_{\rm KS}$ given by the Kolmogorov-Smirnov p-value   for different values of the sampling frequency $f_{\rm s}$  indicated above each panel of the figure. All the histograms were obtained from the same ensemble of $182$ oscillatory hair cells that displayed active oscillations. }
\label{fig:histspval}
\end{figure*}

\begin{figure*}[h!]
\centering
\includegraphics[width=0.9\textwidth]{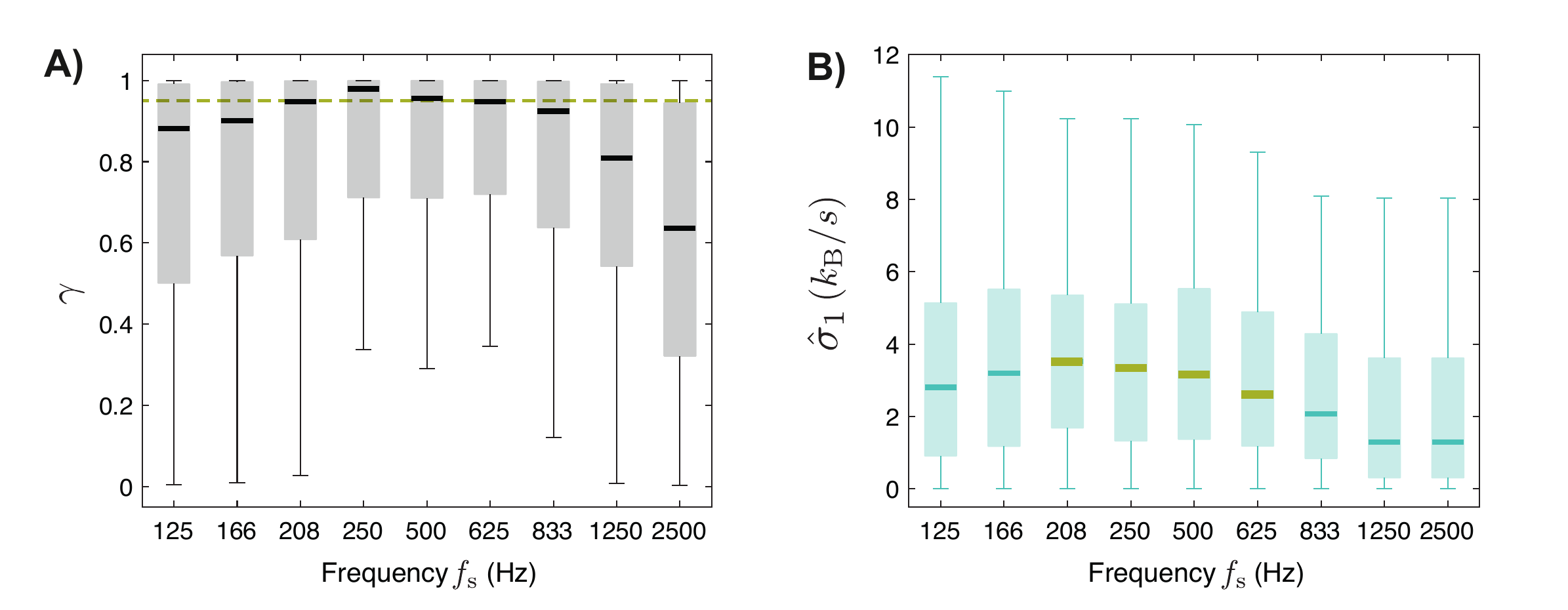}
\caption{Box plots of the parameter $\gamma$ \textbf{(A)} and of the irreversibility measure  $\hat{\sigma}_1$  \textbf{(B)} as a function of the sampling frequency  obtained from recordings  of the tip position of hair bundles in the entire population of 182 cells. The yellow dashed line in (A) is set to the  threshold $\gamma=0.95$ corresponding to the Kolmogorov-Smirnov p-value $p_{\rm KS}<0.05$.  In (B), we highlight (horizontal yellow thick lines) the median of the distributions of $\hat{\sigma}_1$ for which the median value of $\gamma$ is larger than $0.95$. }
\label{fig:power}
\end{figure*}

In Fig.~\ref{fig:histspval} we report the distributions of the parameter  $\gamma$ for the $182$ recordings of spontaneous oscillations at different sampling frequencies.  For all the analyzed cases, the distributions   are right-sweked towards    values of $\gamma$ close to $1$.  Interestingly, the number of cells that  display large $\gamma$  (i.e. KS p-value $p_{\rm KS}<0.05$) attains its maximum in an intermediate frequency range $\sim 200-600$Hz. To gain further insight on this result, we plot in Fig.~\ref{fig:power} box plots of the distributions of  $\gamma$ and of the corresponding irreversibility estimate $\hat{\sigma}_1$ as a function of the data sampling frequency. Notably, the median of  $\gamma$ is above $0.95$  for intermediate sampling frequencies ranging from $208$ to $625$Hz (Fig.~\ref{fig:power}A). For values of $f_s$ of this frequency band $208-625$Hz, the median of $\gamma$  is above $0.95$ indicating that more than half of the cells  display irreversibility with "significant" KS p-value $p_{\rm KS}<0.05$ at those frequencies. Within this band, the median of the irreversibility measure $\hat{\sigma}_1$ decreases monotonically with $f_s$ from    $3.5\,k_{\rm B}/\text{s}$ ($f_{\rm s}=208\,\rm Hz$) to $2.6\,k_{\rm B}/\text{s}$ ($f_{\rm s}=625\,\rm Hz$).

\section{Biophysics of mechanosensory hair bundles}
\label{app:vi}

Details of the experimental procedure have been published elsewhere~\cite{tinevez2007unifying}.  In short, an excised preparation of the bullfrog's (\textit{Rana catesbeiana}) sacculus was mounted on a two-compartment chamber to reproduce the ionic environment of the inner ear.  
This organ is devoted to sensitive detection of low-frequency vibrations ($5-150$ Hz) of the animal's head in a vertical plane; it contains about $3000$ sensory hair cells that are arranged in a planar epithelium.
The basal bodies of hair cells were bathed in a standard saline solution and the hair bundles projected in an artificial endolymph.  The preparation was viewed through a $\times 60$ water-immersion objective of an upright microscope.  Under these conditions, spontaneous hair-bundle oscillations were routinely observed. The oscillations could be recorded by imaging, at a magnification of $\times 1000$, the top of the longest stereociliary row onto a displacement monitor that included a dual photodiode.  Calibration was performed by measuring the output voltages of this photometric system in response to a series of offset displacements.  Here, we analyzed 182 spontaneously oscillating hair bundles from data previously published~\cite{barral2018friction}.

Spontaneous hair-bundle oscillations were described by a published model of active hair-bundle motility~\cite{tinevez2007unifying} that rest  on a necessary condition of negative hair-bundle stiffness, on the presence of molecular motors that actively pull on the tip links, and on feedback by the calcium component of the transduction current.  Hair-bundle deflections affect tension in tip links that interconnect neighbouring stereocillia of the bundle.  Changes in tip-link tension in turn modulate the open probability of mechano-sensitive ion channels connected to these links.  Importantly, the relation between channel gating and tip-link tension is reciprocal: gating of the transduction channels affects tip-link tension.  Consequently, channel gating effectively reduces the stiffness of a hair bundle, a phenomenon appropriately termed "gating compliance",  which can result in negative stiffness if channel-gating forces are strong enough.  Active hair-bundle movements result from the activity of the adaptation motors.  By controlling  tip-link tension, adaptation motors regulate the open probability of the mechanosensitive channels.  The force produced by the motors is in turn regulated by the Ca$^{2+}$ component of the transduction current which thus provides negative feedback on the motor force~\cite{tinevez2007unifying}.   When the fixed point of this dynamical system corresponds to an unstable  position of negative stiffness, the system oscillates spontaneously. The maximal force exerted by the motors $F_{\rm max}$ and the calcium feedback strength $S$ are control parameters of the system and fully determine its dynamics (oscillatory, quiescent, bi-stable)~\cite{nadrowski2004active}.

\section{Quantification of entropy production in numerical simulations of hair bundle oscillations}
\label{app:iv}

In this Section, we provide numerical results for the stochastic model of the ear hair bundle given by Eqs.~(3-5) in the Main Text.  The steady-state entropy production rate of the model is given by
\begin{equation}
\sigma_{\rm tot}=\frac{1}{T}\left\langle  F_1\circ\frac{\text{d}X_1}{\text{d}t}\right\rangle + \frac{1}{T_{\rm eff}} \left\langle F_2 \circ\frac{\text{d}X_2}{\text{d}t}   \right\rangle \quad,\label{eq:s28}
\end{equation}
where $ F_1 = F_1 (X_1,X_2)$, $ F_2 = F_2 (X_1,X_2)$ and $\circ$ denotes the Stratonovich product. Using the definitions of the forces in Eq.~\eqref{eq:s28} one obtains after some algebra Eq.~(6) in the Main Text. 
In all our numerical simulations, we estimate the steady-state averages  of the type
\begin{equation}
\left\langle F \circ \frac{\text{d}X}{\text{d}t}\right\rangle  = \lim_{t\to \infty} \frac{1}{t} \int_0^t F(t') \circ \text{d}X(t')\quad,
\end{equation}
 for a generic force $F(t)=F(X(t),Y(t))$ from a single stationary trajectory of total duration $\tau=300\,\rm s$ and sampling time $\Delta t = 1\,\rm ms$ as follows:
\begin{equation}
\left\langle F \circ \frac{\text{d}X}{\text{d}t}\right\rangle  \simeq \frac{1}{\tau}\sum_{i=1}^{n} \left(\frac{F(t_{i})+F(t_{i-1})}{2}\right) (X(t_{i})-X(t_{i-1}))\quad,
\label{eq:s30}
\end{equation}
where $t_i = i\Delta t$ and $n=\tau/\Delta t$.

\section{Estimation of the quality factor of stochastic oscillations}
\label{app:v}

\begin{figure*}
\centering
\includegraphics[width=\textwidth]{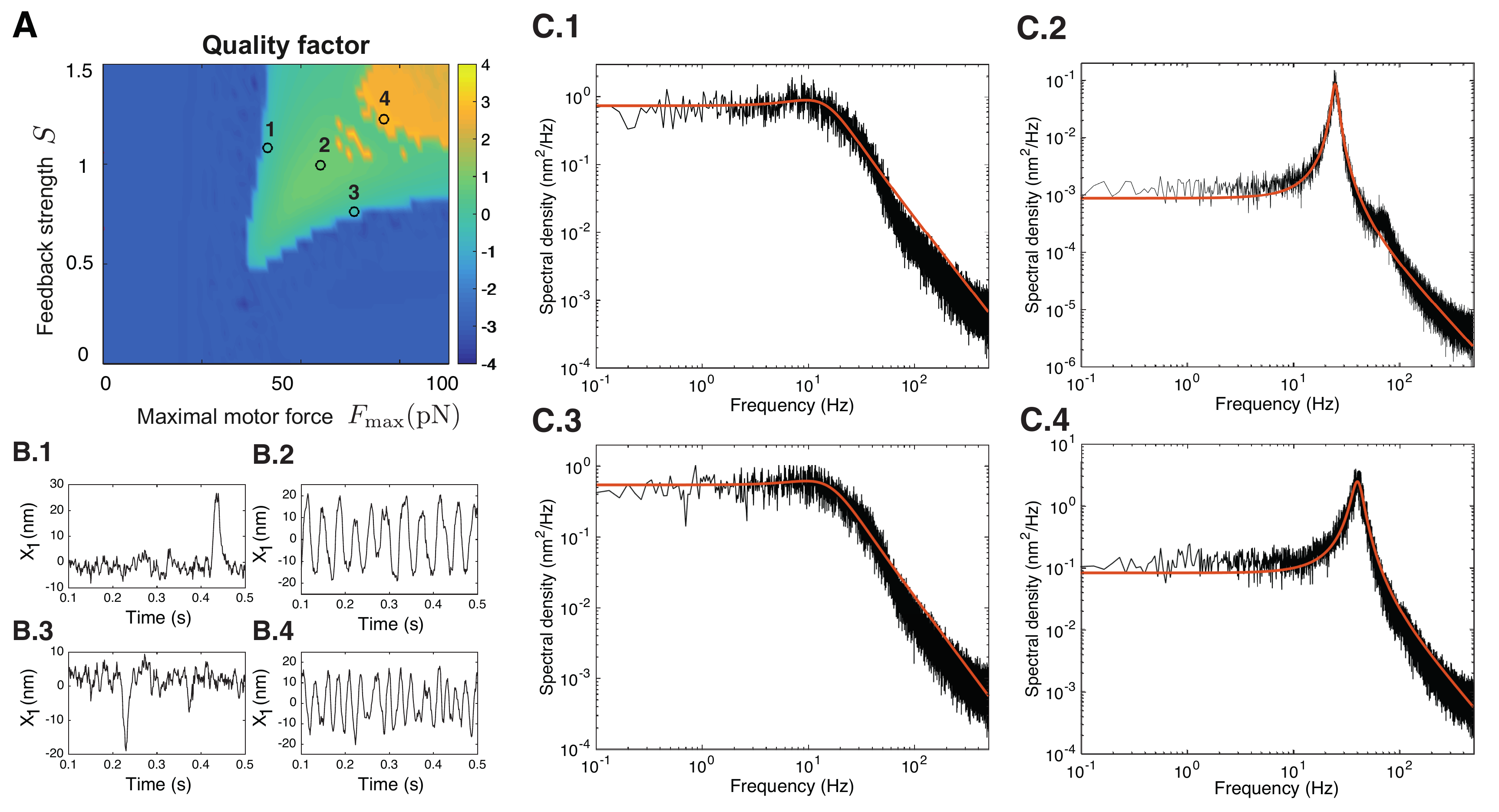}
\caption{ Estimation of the quality factor $Q$ from numerical simulations of the hair bundle.
\textbf{(A)} Values of the quality factor $Q$ calculated from numerical simulations of duration $t_{\rm sim}= 300\,\rm s$  for the same parameter values as in Fig. 4 in the Main Text. 
\textbf{(B)} Examples of $0.5$-second traces of  $X_1$ as a function of time for the parameter values indicated in A: {B.1)} [1 in (A)]; {B.2)} [2 in (A)],  {B.3)} [3 in (A)]; {B.4)} [4 in (A)].  \textbf{(C)} Power spectral density (black line) of the numerical simulations with parameter values indicated with black open circles in (A). The quality factor is estimated from a fit of the power spectra to Eq.~\eqref{eq:PSD} (red line). The values of $Q$ and $f_{\rm o}$ extracted from the fits are: $Q=0.5, f_{\rm o}=7.3\,\text{Hz}$ (C.1), $Q=7, f_{\rm o}=25\,\text{Hz}$ (C.2), $Q=0.45, f_{\rm o}=10.6\,\text{Hz}$ (C.3), $Q=3.8, f_{\rm o}=41.3\,\text{Hz}$ (C.4). 
 \label{fig:siQ}}
\end{figure*}

We estimate the quality factor $Q$ of spontaneous hair-bundle oscillations from numerical simulations of the hair-bundle stochastic model given by Eqs.~(3-4) in the Main Text. For this purpose, we generate a single numerical simulation of duration $t_{\rm sim} = 300\,\rm s$. We then partition the simulation into $10$ consecutive traces of duration $T=t_{\rm sim} /10= 30\,\rm s$. For each of these traces $\{X_{\alpha}(t)\}$ ($\alpha=1,\dots,10$) we compute the power spectral density as $C_\alpha(f) = (1/T) \left| \int_0^T X_\alpha (s) e^{2\pi ift} \,\text{d}t \right|^2$. We then calculate the average of the power spectral density over the $10$ different traces $\tilde{C}(f) =(1/10) \sum_{\alpha=1}^{10}C_\alpha(f)$ and fit the estimate $\tilde{C}(f)$ as a function of $f$  to the sum of two Lorentzian functions~\cite{martin2001comparison,julicher2009spontaneous,barral2018friction}
\begin{equation}
\tilde{C}(f) = \frac{A}{(f_{\rm o}/2Q)^2 + (f-f_{\rm o})^2}+ \frac{A}{(f_{\rm o}/2Q)^2 + (f+f_{\rm o})^2}\quad,
\label{eq:PSD}
\end{equation}
where $Q$ is the quality factor, $f_{\rm o}$ is the oscillation frequency and $A>0$ is an amplitude parameter.  
Figure~\ref{fig:siQ} shows examples of numerical simulations for which we apply this procedure to determine the value of the quality factor by extracting the value $Q$ from the fit of the data to Eq.~\eqref{eq:PSD}. Notably, Eq.~\eqref{eq:PSD} reproduces power spectra of hair-bundle simulations for oscillations with values $Q$ that are in a wide range of orders of magnitude (Fig.~\ref{fig:siQ}C).

\begin{thebibliography}
\expandafter\ifx\csname natexlab\endcsname\relax\def\natexlab#1{#1}\fi
\expandafter\ifx\csname bibnamefont\endcsname\relax
  \def\bibnamefont#1{#1}\fi
\expandafter\ifx\csname bibfnamefont\endcsname\relax
  \def\bibfnamefont#1{#1}\fi
\expandafter\ifx\csname citenamefont\endcsname\relax
  \def\citenamefont#1{#1}\fi
\expandafter\ifx\csname url\endcsname\relax
  \def\url#1{\texttt{#1}}\fi
\expandafter\ifx\csname urlprefix\endcsname\relax\def\urlprefix{URL }\fi
\providecommand{\bibinfo}[2]{#2}
\providecommand{\eprint}[2][]{\url{#2}}

\bibitem[{\citenamefont{Martin et~al.}(2003)\citenamefont{Martin, Bozovic,
  Choe, and Hudspeth}}]{martin2003spontaneous}
\bibinfo{author}{\bibfnamefont{P.}~\bibnamefont{Martin}},
  \bibinfo{author}{\bibfnamefont{D.}~\bibnamefont{Bozovic}},
  \bibinfo{author}{\bibfnamefont{Y.}~\bibnamefont{Choe}}, \bibnamefont{and}
  \bibinfo{author}{\bibfnamefont{A.}~\bibnamefont{Hudspeth}},
  \bibinfo{journal}{J. Neurosci.} \textbf{\bibinfo{volume}{23}},
  \bibinfo{pages}{4533} (\bibinfo{year}{2003}).

\bibitem[{\citenamefont{Tinevez et~al.}(2007)\citenamefont{Tinevez,
  J{\"u}licher, and Martin}}]{tinevez2007unifying}
\bibinfo{author}{\bibfnamefont{J.-Y.} \bibnamefont{Tinevez}},
  \bibinfo{author}{\bibfnamefont{F.}~\bibnamefont{J{\"u}licher}},
  \bibnamefont{and} \bibinfo{author}{\bibfnamefont{P.}~\bibnamefont{Martin}},
  \bibinfo{journal}{Biophys. J.} \textbf{\bibinfo{volume}{93}},
  \bibinfo{pages}{4053} (\bibinfo{year}{2007}).

\bibitem[{\citenamefont{Hudspeth}(2014)}]{hudspeth2014integrating}
\bibinfo{author}{\bibfnamefont{A.}~\bibnamefont{Hudspeth}},
  \bibinfo{journal}{Nature Rev. Neurosci.} \textbf{\bibinfo{volume}{15}},
  \bibinfo{pages}{600} (\bibinfo{year}{2014}).

\bibitem[{\citenamefont{Martin et~al.}(2001)\citenamefont{Martin, Hudspeth, and
  J{\"u}licher}}]{martin2001comparison}
\bibinfo{author}{\bibfnamefont{P.}~\bibnamefont{Martin}},
  \bibinfo{author}{\bibfnamefont{A.}~\bibnamefont{Hudspeth}}, \bibnamefont{and}
  \bibinfo{author}{\bibfnamefont{F.}~\bibnamefont{J{\"u}licher}},
  \bibinfo{journal}{PNAS} \textbf{\bibinfo{volume}{98}}, \bibinfo{pages}{14380}
  (\bibinfo{year}{2001}).

\bibitem[{\citenamefont{Harada and Sasa}(2005)}]{harada2005equality}
\bibinfo{author}{\bibfnamefont{T.}~\bibnamefont{Harada}} \bibnamefont{and}
  \bibinfo{author}{\bibfnamefont{S.-i.} \bibnamefont{Sasa}},
  \bibinfo{journal}{Phys. Rev. Lett.} \textbf{\bibinfo{volume}{95}},
  \bibinfo{pages}{130602} (\bibinfo{year}{2005}).

\bibitem[{\citenamefont{Mizuno et~al.}(2007)\citenamefont{Mizuno, Tardin,
  Schmidt, and MacKintosh}}]{mizuno2007nonequilibrium}
\bibinfo{author}{\bibfnamefont{D.}~\bibnamefont{Mizuno}},
  \bibinfo{author}{\bibfnamefont{C.}~\bibnamefont{Tardin}},
  \bibinfo{author}{\bibfnamefont{C.~F.} \bibnamefont{Schmidt}},
  \bibnamefont{and} \bibinfo{author}{\bibfnamefont{F.~C.}
  \bibnamefont{MacKintosh}}, \bibinfo{journal}{Science}
  \textbf{\bibinfo{volume}{315}}, \bibinfo{pages}{370} (\bibinfo{year}{2007}).

\bibitem[{\citenamefont{Rodr{\'\i}guez-Garc{\'\i}a
  et~al.}(2015)\citenamefont{Rodr{\'\i}guez-Garc{\'\i}a, L{\'o}pez-Montero,
  Mell, Egea, Gov, and Monroy}}]{rodriguez2015direct}
\bibinfo{author}{\bibfnamefont{R.}~\bibnamefont{Rodr{\'\i}guez-Garc{\'\i}a}},
  \bibinfo{author}{\bibfnamefont{I.}~\bibnamefont{L{\'o}pez-Montero}},
  \bibinfo{author}{\bibfnamefont{M.}~\bibnamefont{Mell}},
  \bibinfo{author}{\bibfnamefont{G.}~\bibnamefont{Egea}},
  \bibinfo{author}{\bibfnamefont{N.~S.} \bibnamefont{Gov}}, \bibnamefont{and}
  \bibinfo{author}{\bibfnamefont{F.}~\bibnamefont{Monroy}},
  \bibinfo{journal}{Biophys. J.} \textbf{\bibinfo{volume}{108}},
  \bibinfo{pages}{2794} (\bibinfo{year}{2015}).

\bibitem[{\citenamefont{Turlier et~al.}(2016)\citenamefont{Turlier, Fedosov,
  Audoly, Auth, Gov, Sykes, Joanny, Gompper, and
  Betz}}]{turlier2016equilibrium}
\bibinfo{author}{\bibfnamefont{H.}~\bibnamefont{Turlier}},
  \bibinfo{author}{\bibfnamefont{D.~A.} \bibnamefont{Fedosov}},
  \bibinfo{author}{\bibfnamefont{B.}~\bibnamefont{Audoly}},
  \bibinfo{author}{\bibfnamefont{T.}~\bibnamefont{Auth}},
  \bibinfo{author}{\bibfnamefont{N.~S.} \bibnamefont{Gov}},
  \bibinfo{author}{\bibfnamefont{C.}~\bibnamefont{Sykes}},
  \bibinfo{author}{\bibfnamefont{J.~F.} \bibnamefont{Joanny}},
  \bibinfo{author}{\bibfnamefont{G.}~\bibnamefont{Gompper}}, \bibnamefont{and}
  \bibinfo{author}{\bibfnamefont{T.}~\bibnamefont{Betz}},
  \bibinfo{journal}{Nature Phys.} \textbf{\bibinfo{volume}{12}},
  \bibinfo{pages}{513} (\bibinfo{year}{2016}).

\bibitem[{\citenamefont{Battle et~al.}(2016)\citenamefont{Battle, Broedersz,
  Fakhri, Geyer, Howard, Schmidt, and MacKintosh}}]{battle2016broken}
\bibinfo{author}{\bibfnamefont{C.}~\bibnamefont{Battle}},
  \bibinfo{author}{\bibfnamefont{C.~P.} \bibnamefont{Broedersz}},
  \bibinfo{author}{\bibfnamefont{N.}~\bibnamefont{Fakhri}},
  \bibinfo{author}{\bibfnamefont{V.~F.} \bibnamefont{Geyer}},
  \bibinfo{author}{\bibfnamefont{J.}~\bibnamefont{Howard}},
  \bibinfo{author}{\bibfnamefont{C.~F.} \bibnamefont{Schmidt}},
  \bibnamefont{and} \bibinfo{author}{\bibfnamefont{F.~C.}
  \bibnamefont{MacKintosh}}, \bibinfo{journal}{Science}
  \textbf{\bibinfo{volume}{352}}, \bibinfo{pages}{604} (\bibinfo{year}{2016}).

\bibitem[{\citenamefont{Nardini et~al.}(2017)\citenamefont{Nardini, Fodor,
  Tjhung, Van~Wijland, Tailleur, and Cates}}]{nardini2017entropy}
\bibinfo{author}{\bibfnamefont{C.}~\bibnamefont{Nardini}},
  \bibinfo{author}{\bibfnamefont{{\'E}.}~\bibnamefont{Fodor}},
  \bibinfo{author}{\bibfnamefont{E.}~\bibnamefont{Tjhung}},
  \bibinfo{author}{\bibfnamefont{F.}~\bibnamefont{Van~Wijland}},
  \bibinfo{author}{\bibfnamefont{J.}~\bibnamefont{Tailleur}}, \bibnamefont{and}
  \bibinfo{author}{\bibfnamefont{M.~E.} \bibnamefont{Cates}},
  \bibinfo{journal}{Phys. Rev. X} \textbf{\bibinfo{volume}{7}},
  \bibinfo{pages}{021007} (\bibinfo{year}{2017}).

\bibitem[{\citenamefont{J{\"u}licher et~al.}(1997)\citenamefont{J{\"u}licher,
  Ajdari, and Prost}}]{julicher1997modeling}
\bibinfo{author}{\bibfnamefont{F.}~\bibnamefont{J{\"u}licher}},
  \bibinfo{author}{\bibfnamefont{A.}~\bibnamefont{Ajdari}}, \bibnamefont{and}
  \bibinfo{author}{\bibfnamefont{J.}~\bibnamefont{Prost}},
  \bibinfo{journal}{Rev. Mod. Phys.} \textbf{\bibinfo{volume}{69}},
  \bibinfo{pages}{1269} (\bibinfo{year}{1997}).

\bibitem[{\citenamefont{Keller and
  Bustamante}(2000)}]{keller2000mechanochemistry}
\bibinfo{author}{\bibfnamefont{D.}~\bibnamefont{Keller}} \bibnamefont{and}
  \bibinfo{author}{\bibfnamefont{C.}~\bibnamefont{Bustamante}},
  \bibinfo{journal}{Biophys. J.} \textbf{\bibinfo{volume}{78}},
  \bibinfo{pages}{541} (\bibinfo{year}{2000}).

\bibitem[{\citenamefont{Howard}(2001)}]{howard2001mechanics}
\bibinfo{author}{\bibfnamefont{J.}~\bibnamefont{Howard}},
  \emph{\bibinfo{title}{Mechanics of motor proteins and the cytoskeleton}}
  (\bibinfo{publisher}{Sinauer associates Sunderland, MA},
  \bibinfo{year}{2001}).

\bibitem[{\citenamefont{Qian}(2000)}]{qian2000mathematical}
\bibinfo{author}{\bibfnamefont{H.}~\bibnamefont{Qian}}, \bibinfo{journal}{J.
  Math. Chem.} \textbf{\bibinfo{volume}{27}}, \bibinfo{pages}{219}
  (\bibinfo{year}{2000}).

\bibitem[{\citenamefont{Steinberg}(1986)}]{steinberg1986time}
\bibinfo{author}{\bibfnamefont{I.~Z.} \bibnamefont{Steinberg}},
  \bibinfo{journal}{Biophys. J.} \textbf{\bibinfo{volume}{50}},
  \bibinfo{pages}{171} (\bibinfo{year}{1986}).

\bibitem[{\citenamefont{Hudspeth}(1989)}]{hudspeth1989ear}
\bibinfo{author}{\bibfnamefont{A.~J.} \bibnamefont{Hudspeth}},
  \bibinfo{journal}{Nature} \textbf{\bibinfo{volume}{341}},
  \bibinfo{pages}{397} (\bibinfo{year}{1989}).

\bibitem[{\citenamefont{Martin and Hudspeth}(1999)}]{martin1999active}
\bibinfo{author}{\bibfnamefont{P.}~\bibnamefont{Martin}} \bibnamefont{and}
  \bibinfo{author}{\bibfnamefont{A.}~\bibnamefont{Hudspeth}},
  \bibinfo{journal}{Proceedings of the National Academy of Sciences}
  \textbf{\bibinfo{volume}{96}}, \bibinfo{pages}{14306} (\bibinfo{year}{1999}).

\bibitem[{\citenamefont{Nadrowski et~al.}(2004)\citenamefont{Nadrowski, Martin,
  and J{\"u}licher}}]{nadrowski2004active}
\bibinfo{author}{\bibfnamefont{B.}~\bibnamefont{Nadrowski}},
  \bibinfo{author}{\bibfnamefont{P.}~\bibnamefont{Martin}}, \bibnamefont{and}
  \bibinfo{author}{\bibfnamefont{F.}~\bibnamefont{J{\"u}licher}},
  \bibinfo{journal}{PNAS} \textbf{\bibinfo{volume}{101}},
  \bibinfo{pages}{12195} (\bibinfo{year}{2004}).

\bibitem[{\citenamefont{Le~Goff et~al.}(2005)\citenamefont{Le~Goff, Bozovic,
  and Hudspeth}}]{le2005adaptive}
\bibinfo{author}{\bibfnamefont{L.}~\bibnamefont{Le~Goff}},
  \bibinfo{author}{\bibfnamefont{D.}~\bibnamefont{Bozovic}}, \bibnamefont{and}
  \bibinfo{author}{\bibfnamefont{A.}~\bibnamefont{Hudspeth}},
  \bibinfo{journal}{Proceedings of the National Academy of Sciences}
  \textbf{\bibinfo{volume}{102}}, \bibinfo{pages}{16996}
  (\bibinfo{year}{2005}).

\bibitem[{\citenamefont{Martin and Hudspeth}(2020)}]{martin2020mechanical}
\bibinfo{author}{\bibfnamefont{P.}~\bibnamefont{Martin}} \bibnamefont{and}
  \bibinfo{author}{\bibfnamefont{A.}~\bibnamefont{Hudspeth}},
  \bibinfo{journal}{Annual Review of Condensed Matter Physics}
  \textbf{\bibinfo{volume}{12}} (\bibinfo{year}{2020}).

\bibitem[{\citenamefont{Dinis et~al.}(2012)\citenamefont{Dinis, Martin, Barral,
  Prost, and Joanny}}]{dinis2012fluctuation}
\bibinfo{author}{\bibfnamefont{L.}~\bibnamefont{Dinis}},
  \bibinfo{author}{\bibfnamefont{P.}~\bibnamefont{Martin}},
  \bibinfo{author}{\bibfnamefont{J.}~\bibnamefont{Barral}},
  \bibinfo{author}{\bibfnamefont{J.}~\bibnamefont{Prost}}, \bibnamefont{and}
  \bibinfo{author}{\bibfnamefont{J.~F.} \bibnamefont{Joanny}},
  \bibinfo{journal}{Phys. Rev. Lett.} \textbf{\bibinfo{volume}{109}},
  \bibinfo{pages}{160602} (\bibinfo{year}{2012}).

\bibitem[{\citenamefont{Sekimoto}(2010)}]{sekimoto2010stochastic}
\bibinfo{author}{\bibfnamefont{K.}~\bibnamefont{Sekimoto}},
  \emph{\bibinfo{title}{Stochastic energetics}}, vol. \bibinfo{volume}{799}
  (\bibinfo{publisher}{Springer}, \bibinfo{year}{2010}).

\bibitem[{\citenamefont{Seifert}(2008)}]{seifert2008stochastic}
\bibinfo{author}{\bibfnamefont{U.}~\bibnamefont{Seifert}},
  \bibinfo{journal}{The European Physical Journal B}
  \textbf{\bibinfo{volume}{64}}, \bibinfo{pages}{423} (\bibinfo{year}{2008}).

\bibitem[{\citenamefont{Barral et~al.}(2018)\citenamefont{Barral, J{\"u}licher,
  and Martin}}]{barral2018friction}
\bibinfo{author}{\bibfnamefont{J.}~\bibnamefont{Barral}},
  \bibinfo{author}{\bibfnamefont{F.}~\bibnamefont{J{\"u}licher}},
  \bibnamefont{and} \bibinfo{author}{\bibfnamefont{P.}~\bibnamefont{Martin}},
  \bibinfo{journal}{Biophys. J.} \textbf{\bibinfo{volume}{114}},
  \bibinfo{pages}{425} (\bibinfo{year}{2018}).

\bibitem[{\citenamefont{Barato and Seifert}(2015)}]{barato2015thermodynamic}
\bibinfo{author}{\bibfnamefont{A.~C.} \bibnamefont{Barato}} \bibnamefont{and}
  \bibinfo{author}{\bibfnamefont{U.}~\bibnamefont{Seifert}},
  \bibinfo{journal}{Physical review letters} \textbf{\bibinfo{volume}{114}},
  \bibinfo{pages}{158101} (\bibinfo{year}{2015}).

\bibitem[{\citenamefont{Gingrich et~al.}(2016)\citenamefont{Gingrich, Horowitz,
  Perunov, and England}}]{gingrich2016dissipation}
\bibinfo{author}{\bibfnamefont{T.~R.} \bibnamefont{Gingrich}},
  \bibinfo{author}{\bibfnamefont{J.~M.} \bibnamefont{Horowitz}},
  \bibinfo{author}{\bibfnamefont{N.}~\bibnamefont{Perunov}}, \bibnamefont{and}
  \bibinfo{author}{\bibfnamefont{J.~L.} \bibnamefont{England}},
  \bibinfo{journal}{Physical review letters} \textbf{\bibinfo{volume}{116}},
  \bibinfo{pages}{120601} (\bibinfo{year}{2016}).

\bibitem[{\citenamefont{Lebowitz and Spohn}(1999)}]{lebowitz1999gallavotti}
\bibinfo{author}{\bibfnamefont{J.~L.} \bibnamefont{Lebowitz}} \bibnamefont{and}
  \bibinfo{author}{\bibfnamefont{H.}~\bibnamefont{Spohn}}, \bibinfo{journal}{J.
  Stat. Phys.} \textbf{\bibinfo{volume}{95}}, \bibinfo{pages}{333}
  (\bibinfo{year}{1999}).

\bibitem[{\citenamefont{Maes and Neto{\v{c}}n{\`y}}(2003)}]{maes2003time}
\bibinfo{author}{\bibfnamefont{C.}~\bibnamefont{Maes}} \bibnamefont{and}
  \bibinfo{author}{\bibfnamefont{K.}~\bibnamefont{Neto{\v{c}}n{\`y}}},
  \bibinfo{journal}{J. Stat. Phys.} \textbf{\bibinfo{volume}{110}},
  \bibinfo{pages}{269} (\bibinfo{year}{2003}).

\bibitem[{\citenamefont{Seifert}(2005)}]{seifert2005entropy}
\bibinfo{author}{\bibfnamefont{U.}~\bibnamefont{Seifert}},
  \bibinfo{journal}{Phys. Rev. Lett.} \textbf{\bibinfo{volume}{95}},
  \bibinfo{pages}{040602} (\bibinfo{year}{2005}).

\bibitem[{\citenamefont{Neri et~al.}(2017)\citenamefont{Neri, Rold{\'a}n, and
  J{\"u}licher}}]{neri2017statistics}
\bibinfo{author}{\bibfnamefont{I.}~\bibnamefont{Neri}},
  \bibinfo{author}{\bibfnamefont{{\'E}.}~\bibnamefont{Rold{\'a}n}},
  \bibnamefont{and}
  \bibinfo{author}{\bibfnamefont{F.}~\bibnamefont{J{\"u}licher}},
  \bibinfo{journal}{Phys. Rev. X} \textbf{\bibinfo{volume}{7}},
  \bibinfo{pages}{011019} (\bibinfo{year}{2017}).

\bibitem[{\citenamefont{Gomez-Marin et~al.}(2008)\citenamefont{Gomez-Marin,
  Parrondo, and Van~den Broeck}}]{gomez2008footprints}
\bibinfo{author}{\bibfnamefont{A.}~\bibnamefont{Gomez-Marin}},
  \bibinfo{author}{\bibfnamefont{J.~M.~R.} \bibnamefont{Parrondo}},
  \bibnamefont{and} \bibinfo{author}{\bibfnamefont{C.}~\bibnamefont{Van~den
  Broeck}}, \bibinfo{journal}{EPL} \textbf{\bibinfo{volume}{82}},
  \bibinfo{pages}{50002} (\bibinfo{year}{2008}).

\bibitem[{\citenamefont{Mehl et~al.}(2012)\citenamefont{Mehl, Lander,
  Bechinger, Blickle, and Seifert}}]{mehl2012role}
\bibinfo{author}{\bibfnamefont{J.}~\bibnamefont{Mehl}},
  \bibinfo{author}{\bibfnamefont{B.}~\bibnamefont{Lander}},
  \bibinfo{author}{\bibfnamefont{C.}~\bibnamefont{Bechinger}},
  \bibinfo{author}{\bibfnamefont{V.}~\bibnamefont{Blickle}}, \bibnamefont{and}
  \bibinfo{author}{\bibfnamefont{U.}~\bibnamefont{Seifert}},
  \bibinfo{journal}{Phys. Rev. Lett.} \textbf{\bibinfo{volume}{108}},
  \bibinfo{pages}{220601} (\bibinfo{year}{2012}).

\bibitem[{\citenamefont{Celani et~al.}(2012)\citenamefont{Celani, Bo, Eichhorn,
  and Aurell}}]{celani2012anomalous}
\bibinfo{author}{\bibfnamefont{A.}~\bibnamefont{Celani}},
  \bibinfo{author}{\bibfnamefont{S.}~\bibnamefont{Bo}},
  \bibinfo{author}{\bibfnamefont{R.}~\bibnamefont{Eichhorn}}, \bibnamefont{and}
  \bibinfo{author}{\bibfnamefont{E.}~\bibnamefont{Aurell}},
  \bibinfo{journal}{Phys. Rev. Lett.} \textbf{\bibinfo{volume}{109}},
  \bibinfo{pages}{260603} (\bibinfo{year}{2012}).

\bibitem[{\citenamefont{Rold{\'a}n and Parrondo}(2010)}]{roldan2010estimating}
\bibinfo{author}{\bibfnamefont{E.}~\bibnamefont{Rold{\'a}n}} \bibnamefont{and}
  \bibinfo{author}{\bibfnamefont{J.~M.~R.} \bibnamefont{Parrondo}},
  \bibinfo{journal}{Phys. Rev. Lett.} \textbf{\bibinfo{volume}{105}},
  \bibinfo{pages}{150607} (\bibinfo{year}{2010}).

\bibitem[{\citenamefont{Rold{\'a}n and Parrondo}(2012)}]{roldan2012entropy}
\bibinfo{author}{\bibfnamefont{{\'E}.}~\bibnamefont{Rold{\'a}n}}
  \bibnamefont{and} \bibinfo{author}{\bibfnamefont{J.~M.~R.}
  \bibnamefont{Parrondo}}, \bibinfo{journal}{Phys. Rev. E}
  \textbf{\bibinfo{volume}{85}}, \bibinfo{pages}{031129}
  (\bibinfo{year}{2012}).

\bibitem[{\citenamefont{Andrieux et~al.}(2008)\citenamefont{Andrieux, Gaspard,
  Ciliberto, Garnier, Joubaud, and Petrosyan}}]{andrieux2008thermodynamic}
\bibinfo{author}{\bibfnamefont{D.}~\bibnamefont{Andrieux}},
  \bibinfo{author}{\bibfnamefont{P.}~\bibnamefont{Gaspard}},
  \bibinfo{author}{\bibfnamefont{S.}~\bibnamefont{Ciliberto}},
  \bibinfo{author}{\bibfnamefont{N.}~\bibnamefont{Garnier}},
  \bibinfo{author}{\bibfnamefont{S.}~\bibnamefont{Joubaud}}, \bibnamefont{and}
  \bibinfo{author}{\bibfnamefont{A.}~\bibnamefont{Petrosyan}},
  \bibinfo{journal}{J. Stat. Mech.} \textbf{\bibinfo{volume}{2008}},
  \bibinfo{pages}{P01002} (\bibinfo{year}{2008}).

\bibitem[{\citenamefont{Tusch et~al.}(2014)\citenamefont{Tusch, Kundu, Verley,
  Blondel, Miralles, D{\'e}moulin, Lacoste, and Baudry}}]{tusch2014energy}
\bibinfo{author}{\bibfnamefont{S.}~\bibnamefont{Tusch}},
  \bibinfo{author}{\bibfnamefont{A.}~\bibnamefont{Kundu}},
  \bibinfo{author}{\bibfnamefont{G.}~\bibnamefont{Verley}},
  \bibinfo{author}{\bibfnamefont{T.}~\bibnamefont{Blondel}},
  \bibinfo{author}{\bibfnamefont{V.}~\bibnamefont{Miralles}},
  \bibinfo{author}{\bibfnamefont{D.}~\bibnamefont{D{\'e}moulin}},
  \bibinfo{author}{\bibfnamefont{D.}~\bibnamefont{Lacoste}}, \bibnamefont{and}
  \bibinfo{author}{\bibfnamefont{J.}~\bibnamefont{Baudry}},
  \bibinfo{journal}{Phys. Rev. Lett.} \textbf{\bibinfo{volume}{112}},
  \bibinfo{pages}{180604} (\bibinfo{year}{2014}).

\bibitem[{\citenamefont{Rold\'an}(2014)}]{roldan2014irreversibility}
\bibinfo{author}{\bibfnamefont{E.}~\bibnamefont{Rold\'an}},
  \emph{\bibinfo{title}{Irreversibility and dissipation in microscopic
  systems}} (\bibinfo{publisher}{Springer Theses}, \bibinfo{address}{Berlin},
  \bibinfo{year}{2014}).

\bibitem[{\citenamefont{Efron and Jeen}(1994)}]{efron1994detection}
\bibinfo{author}{\bibfnamefont{A.~J.} \bibnamefont{Efron}} \bibnamefont{and}
  \bibinfo{author}{\bibfnamefont{H.}~\bibnamefont{Jeen}},
  \bibinfo{journal}{IEEE Trans. Sign. Proc.} \textbf{\bibinfo{volume}{42}},
  \bibinfo{pages}{1572} (\bibinfo{year}{1994}).

\bibitem[{\citenamefont{Galka et~al.}(2006)\citenamefont{Galka, Ozaki, Bayard,
  and Yamashita}}]{galka2006whitening}
\bibinfo{author}{\bibfnamefont{A.}~\bibnamefont{Galka}},
  \bibinfo{author}{\bibfnamefont{T.}~\bibnamefont{Ozaki}},
  \bibinfo{author}{\bibfnamefont{J.~B.} \bibnamefont{Bayard}},
  \bibnamefont{and}
  \bibinfo{author}{\bibfnamefont{O.}~\bibnamefont{Yamashita}},
  \bibinfo{journal}{J. Stat. Phys.} \textbf{\bibinfo{volume}{124}},
  \bibinfo{pages}{1275} (\bibinfo{year}{2006}).

\bibitem[{\citenamefont{Bonachela et~al.}(2008)\citenamefont{Bonachela,
  Hinrichsen, and Munoz}}]{bonachela2008entropy}
\bibinfo{author}{\bibfnamefont{J.~A.} \bibnamefont{Bonachela}},
  \bibinfo{author}{\bibfnamefont{H.}~\bibnamefont{Hinrichsen}},
  \bibnamefont{and} \bibinfo{author}{\bibfnamefont{M.~A.} \bibnamefont{Munoz}},
  \bibinfo{journal}{Journal of Physics A: Mathematical and Theoretical}
  \textbf{\bibinfo{volume}{41}}, \bibinfo{pages}{202001}
  (\bibinfo{year}{2008}).

\bibitem[{\citenamefont{Gammaitoni et~al.}(1998)\citenamefont{Gammaitoni,
  H{\"a}nggi, Jung, and Marchesoni}}]{gammaitoni1998stochastic}
\bibinfo{author}{\bibfnamefont{L.}~\bibnamefont{Gammaitoni}},
  \bibinfo{author}{\bibfnamefont{P.}~\bibnamefont{H{\"a}nggi}},
  \bibinfo{author}{\bibfnamefont{P.}~\bibnamefont{Jung}}, \bibnamefont{and}
  \bibinfo{author}{\bibfnamefont{F.}~\bibnamefont{Marchesoni}},
  \bibinfo{journal}{Reviews of modern physics} \textbf{\bibinfo{volume}{70}},
  \bibinfo{pages}{223} (\bibinfo{year}{1998}).

\bibitem[{\citenamefont{\text{See Supplemental Material}}()}]{si}
\bibinfo{author}{\bibnamefont{\text{See Supplemental Material}}}.

\bibitem[{\citenamefont{Shiraishi et~al.}(2016)\citenamefont{Shiraishi, Saito,
  and Tasaki}}]{shiraishi2016universal}
\bibinfo{author}{\bibfnamefont{N.}~\bibnamefont{Shiraishi}},
  \bibinfo{author}{\bibfnamefont{K.}~\bibnamefont{Saito}}, \bibnamefont{and}
  \bibinfo{author}{\bibfnamefont{H.}~\bibnamefont{Tasaki}},
  \bibinfo{journal}{Phys. Rev. Lett.} \textbf{\bibinfo{volume}{117}},
  \bibinfo{pages}{190601} (\bibinfo{year}{2016}).

\bibitem[{\citenamefont{Bormuth et~al.}(2014)\citenamefont{Bormuth, Barral,
  Joanny, J{\"u}licher, and Martin}}]{bormuth2014transduction}
\bibinfo{author}{\bibfnamefont{V.}~\bibnamefont{Bormuth}},
  \bibinfo{author}{\bibfnamefont{J.}~\bibnamefont{Barral}},
  \bibinfo{author}{\bibfnamefont{J.-F.} \bibnamefont{Joanny}},
  \bibinfo{author}{\bibfnamefont{F.}~\bibnamefont{J{\"u}licher}},
  \bibnamefont{and} \bibinfo{author}{\bibfnamefont{P.}~\bibnamefont{Martin}},
  \bibinfo{journal}{PNAS} \textbf{\bibinfo{volume}{111}}, \bibinfo{pages}{7185}
  (\bibinfo{year}{2014}).

\bibitem[{\citenamefont{Chetrite and Gawedzki}(2008)}]{chetrite2008fluctuation}
\bibinfo{author}{\bibfnamefont{R.}~\bibnamefont{Chetrite}} \bibnamefont{and}
  \bibinfo{author}{\bibfnamefont{K.}~\bibnamefont{Gawedzki}},
  \bibinfo{journal}{Comm. Math. Phys.} \textbf{\bibinfo{volume}{282}},
  \bibinfo{pages}{469} (\bibinfo{year}{2008}).

\bibitem[{\citenamefont{Dabelow et~al.}(2018)\citenamefont{Dabelow, Bo, and
  Eichhorn}}]{dabelow2018irreversibility}
\bibinfo{author}{\bibfnamefont{L.}~\bibnamefont{Dabelow}},
  \bibinfo{author}{\bibfnamefont{S.}~\bibnamefont{Bo}}, \bibnamefont{and}
  \bibinfo{author}{\bibfnamefont{R.}~\bibnamefont{Eichhorn}},
  \bibinfo{journal}{arXiv:1806.04956}  (\bibinfo{year}{2018}).

\bibitem[{\citenamefont{Sekimoto}(1998)}]{sekimoto1998langevin}
\bibinfo{author}{\bibfnamefont{K.}~\bibnamefont{Sekimoto}},
  \bibinfo{journal}{Prog. Theor. Phys. Suppl.} \textbf{\bibinfo{volume}{130}},
  \bibinfo{pages}{17} (\bibinfo{year}{1998}).

\bibitem[{\citenamefont{Rold{\'a}n et~al.}(2015)\citenamefont{Rold{\'a}n, Neri,
  D{\"o}rpinghaus, Meyr, and J{\"u}licher}}]{roldan2015decision}
\bibinfo{author}{\bibfnamefont{{\'E}.}~\bibnamefont{Rold{\'a}n}},
  \bibinfo{author}{\bibfnamefont{I.}~\bibnamefont{Neri}},
  \bibinfo{author}{\bibfnamefont{M.}~\bibnamefont{D{\"o}rpinghaus}},
  \bibinfo{author}{\bibfnamefont{H.}~\bibnamefont{Meyr}}, \bibnamefont{and}
  \bibinfo{author}{\bibfnamefont{F.}~\bibnamefont{J{\"u}licher}},
  \bibinfo{journal}{Phys. Rev. Lett.} \textbf{\bibinfo{volume}{115}},
  \bibinfo{pages}{250602} (\bibinfo{year}{2015}).

\bibitem[{\citenamefont{Pietzonka et~al.}(2016)\citenamefont{Pietzonka, Barato,
  and Seifert}}]{pietzonka2016universal}
\bibinfo{author}{\bibfnamefont{P.}~\bibnamefont{Pietzonka}},
  \bibinfo{author}{\bibfnamefont{A.~C.} \bibnamefont{Barato}},
  \bibnamefont{and} \bibinfo{author}{\bibfnamefont{U.}~\bibnamefont{Seifert}},
  \bibinfo{journal}{Phys. Rev. E} \textbf{\bibinfo{volume}{93}},
  \bibinfo{pages}{052145} (\bibinfo{year}{2016}).

\bibitem[{\citenamefont{Maes}(2017)}]{maes2017frenetic}
\bibinfo{author}{\bibfnamefont{C.}~\bibnamefont{Maes}}, \bibinfo{journal}{Phys.
  Rev. Lett.} \textbf{\bibinfo{volume}{119}}, \bibinfo{pages}{160601}
  (\bibinfo{year}{2017}).

\bibitem[{\citenamefont{Li et~al.}(2019)\citenamefont{Li, Horowitz, Gingrich,
  and Fakhri}}]{li2019quantifying}
\bibinfo{author}{\bibfnamefont{J.}~\bibnamefont{Li}},
  \bibinfo{author}{\bibfnamefont{J.~M.} \bibnamefont{Horowitz}},
  \bibinfo{author}{\bibfnamefont{T.~R.} \bibnamefont{Gingrich}},
  \bibnamefont{and} \bibinfo{author}{\bibfnamefont{N.}~\bibnamefont{Fakhri}},
  \bibinfo{journal}{Nature communications} \textbf{\bibinfo{volume}{10}},
  \bibinfo{pages}{1} (\bibinfo{year}{2019}).

\bibitem[{\citenamefont{Frishman and Ronceray}(2018)}]{frishman2018learning}
\bibinfo{author}{\bibfnamefont{A.}~\bibnamefont{Frishman}} \bibnamefont{and}
  \bibinfo{author}{\bibfnamefont{P.}~\bibnamefont{Ronceray}},
  \bibinfo{journal}{arXiv:1809.09650}  (\bibinfo{year}{2018}).

\bibitem[{\citenamefont{Van~Vu et~al.}(2020)\citenamefont{Van~Vu, Hasegawa
  et~al.}}]{van2020unified}
\bibinfo{author}{\bibfnamefont{T.}~\bibnamefont{Van~Vu}},
  \bibinfo{author}{\bibfnamefont{Y.}~\bibnamefont{Hasegawa}},
  \bibnamefont{et~al.}, \bibinfo{journal}{Physical Review E}
  \textbf{\bibinfo{volume}{102}}, \bibinfo{pages}{062132}
  (\bibinfo{year}{2020}).

\bibitem[{\citenamefont{Pietzonka et~al.}(2017)\citenamefont{Pietzonka, Ritort,
  and Seifert}}]{pietzonka2017finite}
\bibinfo{author}{\bibfnamefont{P.}~\bibnamefont{Pietzonka}},
  \bibinfo{author}{\bibfnamefont{F.}~\bibnamefont{Ritort}}, \bibnamefont{and}
  \bibinfo{author}{\bibfnamefont{U.}~\bibnamefont{Seifert}},
  \bibinfo{journal}{Physical Review E} \textbf{\bibinfo{volume}{96}},
  \bibinfo{pages}{012101} (\bibinfo{year}{2017}).

\bibitem[{\citenamefont{Seara et~al.}(2021)\citenamefont{Seara, Machta, and
  Murrell}}]{seara2021irreversibility}
\bibinfo{author}{\bibfnamefont{D.~S.} \bibnamefont{Seara}},
  \bibinfo{author}{\bibfnamefont{B.~B.} \bibnamefont{Machta}},
  \bibnamefont{and} \bibinfo{author}{\bibfnamefont{M.~P.}
  \bibnamefont{Murrell}}, \bibinfo{journal}{Nature Communications}
  \textbf{\bibinfo{volume}{12}}, \bibinfo{pages}{1} (\bibinfo{year}{2021}).

\bibitem[{\citenamefont{J{\"u}licher et~al.}(2009)\citenamefont{J{\"u}licher,
  Dierkes, Lindner, Prost, and Martin}}]{julicher2009spontaneous}
\bibinfo{author}{\bibfnamefont{F.}~\bibnamefont{J{\"u}licher}},
  \bibinfo{author}{\bibfnamefont{K.}~\bibnamefont{Dierkes}},
  \bibinfo{author}{\bibfnamefont{B.}~\bibnamefont{Lindner}},
  \bibinfo{author}{\bibfnamefont{J.}~\bibnamefont{Prost}}, \bibnamefont{and}
  \bibinfo{author}{\bibfnamefont{P.}~\bibnamefont{Martin}},
  \bibinfo{journal}{Eur. Phys. J. E} \textbf{\bibinfo{volume}{29}},
  \bibinfo{pages}{449} (\bibinfo{year}{2009}).

\end{thebibliography}
\end{document}